# Energy-Aware Earth System Science: A Community Roadmap

Colfescu, I.[1,2]; Bracco, A.[3]; Moraw, K.[4]; Lu, C.[1]; Bareford, M.[4]; Beucler, T.[5]; Bollasina, M.[6]; Bilbao-Ludena, J. C.[7]; Christensen, H.[8]; DiGioia, S.[9]; Hakim, G.[10]; Hauser, T.[11]; Jackson, A.[4]; Le, P.[12]; Manco, I.[3]; Mobbs, S.[2]; Mozaffari, A.[13]; Navarra, A.[3]; Parsons, M.[4]; Rae, J. W. B.[14]; Reichler, T.[15]; Rodriguez-Pardo, C.[3,16,17]; Stacey, G.[18]; Weiland, M.[4]

[1]University of St Andrews / National Centre for Atmospheric Science (NCAS), UK
[2]NCAS, UK
[3]CMCC Foundation, Euro-Mediterranean Center on Climate Change, Italy
[4]EPCC, University of Edinburgh, UK
[5]University of Lausanne, Switzerland
[6]University of Edinburgh, UK
[7]NCAS, Department of Meteorology, University of Reading, UK
[8]University of Oxford, UK
[9]ICTP/SISSA, Italy
[10]University of Washington, USA
[11]University of Colorado Boulder, Boulder, CO, USA
[12]School of Computer Science, University of St Andrews, UK
[13]Barcelona Supercomputing Centre, Spain
[14]School of Earth and Environmental Sciences, University of St Andrews, UK
[15]Department of Atmospheric Sciences, University of Utah, Salt Lake City, USA
[16]Politecnico di Milano, Italy
[17]RFF/CMCC European Institute on Economics and the Environment (EIEE), Italy
[18]European Leadership Network (ELN), UK

Corresponding author: ioana.colfescu@ncas.ac.uk

## Abstract (150 words)

Earth system models underpin the evidence that informs net-zero policy, yet their energy demands are rarely considered when judging model quality. Using archived output from 159 models spanning five generations of the Coupled Model Intercomparison Project (CMIP), we show that model error has fallen at a constant rate of 13.5 per cent per doubling of compute since 1995. Halving CMIP6 error would require 28 times more compute. Improvements are uneven, with sea-level pressure and surface air temperature errors declining 3.5-fold and 2.7-fold since CMIP1, but precipitation error by only 7 per cent. Machine-learning weather models match or exceed operational forecast skill at a fraction of the runtime, but none publishes measured energy, and including training changes how they rank. We propose a skill–energy framework that combines an empirical Pareto front with a ranking score. These findings argue for standardised reporting of energy use and performance ahead of CMIP8.

## Motivation: From the Charney Ladder to the Skill–Energy Frontier

Earth system models support the scientific evidence that informs climate policy, yet their energy demands are rarely reported or assessed against predictive skill. Since 1995, the Coupled Model Intercomparison Project (CMIP), coordinated by the World Climate Research Programme (WCRP), has enabled systematic, community-wide assessment of these models (Meehl et al., 2000; Eyring et al., 2016). This coordination has supported substantial progress. Models capture many large-scale trends and regional responses more accurately than before, and projections of key phenomena show greater agreement across modelling groups (Xin et al., 2020; Moon et al., 2022; Weijer et al., 2020; McCarthy et al., 2023). Successive methodological advances have expanded model capabilities but also raised computational demands (Manabe and Stouffer, 1988; Molteni et al., 1996; Palmer et al., 2019), and the tens of petabytes produced by CMIP6 strain not just compute but also archives and analysis workflows (Cimadevilla et al., 2024). Yet, systematic biases persist across the 21 endorsed Model Intercomparison Projects (MIPs) of CMIP6 despite sharply increasing costs (Juckes et al., 2020).

The Charney Ladder (Balaji, 2021) has been proposed to describe the progression towards greater model complexity and finer resolution, concurrently with exponentially increasing computing resources. This paper builds upon the Charney Ladder and asks whether scientific return has kept pace. Growing demands for compute, storage and data movement place pressure on research budgets and commitments to reduce emissions (Broude et al., 2021; Khandelwal et al., 2024). At exascale, energy availability and data movement constrain weather and climate simulations, rather than arithmetic capability alone (Schulthess et al., 2019; Neumann et al., 2019). Figure 1a shows the growth in estimated computational demand, derived from each model's native resolution rather than from the growth in hardware capability tracked by Balaji (2021).

Here we examined how model skill has changed with estimated computational demand. We recomputed the $I^2$ error index of Reichler and Kim (2008) using archived output from 159 models spanning five CMIP generations. The index is a normalised measure of error, and lower values indicate closer agreement with observations. The median error index falls from 1.56 in CMIP1 to 0.61 in CMIP6 (Fig. 1b), while resolution-based estimates of compute increase about 29-fold for a typical model (generation geometric means, Fig. 1a). The reductions between successive generations, 21, 28, 21 and 14 per cent, are irregular. A more consistent pattern emerges when these reductions are expressed per doubling of estimated compute. Across the three intergenerational transitions for which the comparison is well posed, error decreases by 13.5 per cent per doubling, with a standard deviation of 0.7 percentage points (Supplementary Section S1 and Table S1). Meanwhile, the increase in compute between successive generations has slowed, from 2.2 doublings between CMIP2 and CMIP3 to 1.1 between CMIP5 and CMIP6. Extrapolating the error–compute relationship suggests that halving CMIP6 error would require about 28 times its compute. At 13.5 per cent per doubling, halving takes about 4.8 doublings, because 0.865 to the power 4.8 is 0.5, and 4.8 doublings multiply compute by about 28. Compute is not the only driver of these gains. More realistic formulation and parameterisation of physical processes also improve skill, and across generations the two cannot be separated. The rate per doubling therefore measures the joint return on compute and scientific development. These resolution-based estimates describe computational demand, not energy consumption, and do not capture all differences in model complexity or implementation (Supplementary Fig. S1). Expensive models are also costly to tune, because each tuning experiment costs as much as the model. The most expensive models are therefore likely to fall short of the skill their resolution would allow.

A single index also conceals substantial differences between variables. Between CMIP1 and CMIP6, sea-level pressure error fell 3.5-fold, and surface air temperature error fell 2.7-fold, whereas precipitation error decreased by only 7 per cent. Despite three decades of added resolution and complexity, improvement in precipitation remains small under this metric, although precipitation is central to climate-impact assessments. We therefore examine where the return occurs as well as its total. Supplementary Section S1 and Tables S4 and S5 provide the full decomposition.

The energy footprint of Earth system models remains largely unquantified. The Computational Performance Model Intercomparison Project (CPMIP; Balaji et al., 2017) established a framework for collecting standardised computational performance and energy metrics, with more recent reporting beginning to address this gap (Acosta et al., 2024). In 2024, the WCRP established a CMIP7 Energy and Carbon Footprint Task Team (WCRP, 2024). The eight institutions that reported complete data for CMIP6 account for an operational footprint of 1,692 tonnes of CO2. This is about one hundred-thousandth of the 180 million tonnes of $CO_2$ emitted each year by data centres worldwide (IEA, 2025). In a separate kilometre-scale case study, including emissions embodied in hardware increased the total footprint by a factor of approximately 1.6 (Palomas et al., 2026). These findings show how scarce reporting is and why accounting boundaries matter. The wider context adds urgency. Global electricity demand is projected to increase significantly by 2030, with artificial intelligence (AI) a major driver alongside electric vehicles, industrialisation and cooling (IEA, 2025).

AI offers a route to improving computational efficiency in Earth system science (Bauer et al., 2021). Balaji (2021) argues that, as gains from faster chips slow with the end of Dennard scaling, machine learning offers a way back down the Charney Ladder (see also Balaji et al., 2022). Weather forecasting underwent a "quiet revolution" of sustained improvement over half a century (Bauer et al., 2015), but from around 2022, AI-based systems demonstrated competitive forecast skill with potentially lower energy requirements (Lam et al., 2023; Bi et al., 2023; Pathak et al., 2022; Chen et al., 2023a, b). Their predictive skill is well documented. For 500 hPa geopotential height at a five-day lead time, the comparison in Fig. 1c, based on WeatherBench 2 (Rasp et al., 2024), shows GraphCast (Lam et al., 2023), Pangu-Weather (Bi et al., 2023), FuXi (Chen et al., 2023b) and GenCast (Price et al., 2025) matching or exceeding the operational high-resolution forecast of the European Centre for Medium-Range Weather Forecasts (IFS HRES).

Evidence of the corresponding energy advantage is less conclusive. Among the weather prediction systems assessed here, we found no published measurements of forecast-inference energy: the energy required to run a trained model to produce a forecast. Several models also report no training compute (Mozaffari et al., 2026). Where inference time and hardware are reported, reconstruction suggests energy requirements several orders of magnitude below those of a conventional forecast cycle. Such comparisons exclude training, architecture search and production of the reanalysis data on which these models depend. An estimate assembled from published training and inference figures suggests that data-driven models use 21 to 1,273 times less energy than a physics-based model over a year of operation, including training (Rieutord et al., 2026).

Transferring these advances from weather forecasting to climate projection presents several challenges. Climate projections must represent responses to changing external forcing, including climate states outside the observational training record, while capturing interactions among the atmosphere, ocean, land surface and cryosphere over long timescales. Progress towards AI-based ESMs has therefore been slower than progress in weather forecasting (Bracco et al., 2025; Antonio et al., 2026). Hybrid models increasingly incorporate AI components while retaining physical constraints to improve reliability and consistency (Eyring et al., 2024; Grundner et al., 2025; Kochkov et al., 2024). Constraints based on conservation laws and equations of state can improve data efficiency and may reduce the energy required for data generation, training and inference, although these benefits require evaluation for each application. Climate-invariant methods seek to improve generalisation beyond the training data by preserving physically relevant relationships under changes in climate (Beucler et al., 2024). These developments broaden the range of approaches that a skill–energy framework must accommodate. Defining explicit and comparable accounting boundaries for such a framework is the central motivation of this work.

We propose the Skill–Energy Frontier, an extension of the Charney Ladder that evaluates predictive skill alongside energy demand. The framework compares both weather and climate prediction systems within defined task classes, here medium-range weather forecasting and climate simulation. It uses an empirical Pareto front, the set of systems that no other system beats on both skill and energy, as the diagnostic and a complementary score for ranking. We also introduce a five-level readiness scale for energy reporting, apply it to the CMIP6 community, and identify four pillars for improving skill–energy performance. Together, these contributions provide a basis for assessing whether methodological advances improve skill, reduce energy demand or achieve both.

## The Skill–Energy plane and Energy-Adjusted Skill Score (EASS)

Our framework builds on earlier calls for sustainable computational science, including the GREENER principles (Lannelongue et al., 2023), studies of the carbon footprint of scientific computing (Henderson et al., 2020; Patterson et al., 2022), and proposals to report efficiency alongside accuracy in machine learning (Schwartz et al., 2020).

*Nature Computational Science* has asked authors to estimate and report the carbon footprint of their computations (Nature Computational Science, 2023). A recent checklist for machine learning in Earth system modelling recommends tracking energy and carbon throughout the modelling workflow and proposes an energy-to-accuracy ratio for comparing configurations (Dainelli et al., 2026). We adapt these approaches to Earth system prediction, following the directions identified for next-generation multiscale modelling (Eyring et al., 2024; Bracco et al., 2025).

Existing benchmarks address predictive skill or computational cost separately, and none integrates energy accounting into its skill assessment (Supplementary Section S4.1). CPMIP reports computational performance and energy for CMIP6 models but does not assess predictive skill (Acosta et al., 2024). The Skill–Energy plane shows both: the skill a system delivers and the energy it uses.

The Skill–Energy plane places each system as a point on two axes (Box 1; Figs. 1c, 2a). The first is energy, E, measured in kilowatt-hours (kWh), accounting for the operational workflow: data preparation, initialisation, production simulation or inference, post-processing and storage attributable to the reported forecast cycle or simulation period. It excludes model development. The second axis is skill where $\xi$ is a dimensionless error ratio that compares a system's error with that of a naive reference. For a weather forecast, $\xi$ is the mean squared error divided by that of climatology, the long-term average for that time of year. For a climate simulation, $\xi$ is the Reichler & Kim index $I^2$: the error of the simulated climatology divided by the observed year-to-year variance, so that $\chi_h = 1/(1 + I^2)$ (Figs. 1b, 2). A perfect prediction scores 1. An error ratio of 1, meaning the system does no better than the reference, gives 0.5. Supplementary Section S2 gives the derivation, and Table 1 lists the variables, evaluation periods and reference definitions. With energy on the horizontal axis and skill on the vertical axis, progress towards the upper left means higher skill at lower energy demand (Fig. 1c). Whereas the Charney Ladder describes advances in capability supported by increasing computational resources, this framework also recognises advances that reduce the energy required to achieve a given level of skill.

The main diagnostic is the empirical Pareto front (Beucler et al., 2025). The front is the staircase boundary formed by systems with the best combination of skill and energy use (Figs. 1c, 2a). A system is empirically Pareto-optimal within its comparison class (weather or climate here) when no other system in that class is both cheaper and more skilful. The true front, set by the joint limits of predictability and efficiency, is unknown. The empirical front reflects only the systems assessed. We recommend reporting it because a single number cannot show the trade-off between skill and energy.

For applications requiring a ranking, we define the Energy-Adjusted Skill Score:

EASS expresses skill per unit of energy and has units of $kWh^{-1}$ (Figs. 1d, 2c). It complements the Pareto front. Across the systems considered here, reported energy spans about seven orders of magnitude, whereas skill varies much less ($\chi_h$ from 0.49 to 0.91, Table 1 and Fig. 2). EASS therefore largely reflects energy differences, while skill mainly distinguishes systems with similar energy use. A system can obtain a high EASS through very low energy use despite weak predictive performance. EASS should consequently be reported alongside its underlying skill and energy values and should not be used as a standalone optimisation target.

The energy assigned to a system, and therefore its position on the plane and its EASS, depends on which costs are included. We distinguish operational, lifecycle and system-level accounting, and reports must state which components are measured, estimated or unavailable (Supplementary Section S4.2). For AI systems, excluding training and other development costs can substantially overstate efficiency, so AI and hybrid systems should report operational and lifecycle EASS separately.

Normalising skill does not make different prediction tasks interchangeable. Skill metrics, reference definitions, physical content and production units differ between task classes, so we compare systems only within a class and under a common evaluation protocol. Medium-range numerical weather prediction (NWP) is assessed at a fixed lead time and costed per forecast cycle. Decadal prediction and longer projections are assessed using climatological means and costed per simulated year. Figures 1c, d and 2 present the two classes separately, and Table 1 identifies the class of each system. Differences in forecast coverage, resolution, ensemble size and other outputs must also be stated, because a shared skill metric may not capture all capabilities delivered by the reported energy expenditure.

Table 1 reports EASS for both classes using the framework summarised in Box 1, while distinguishing measured from reconstructed energy. Climate-model energy data come from CPMIP (Acosta et al., 2024) for the thirteen CMIP6 models that can be matched to models scored here (Fig. 2b, c). No equivalent energy data are available in our assessment for earlier CMIP generations. Weather-system energy is reconstructed rather than measured. Four of the six systems report both inference time and hardware, permitting an approximate reconstruction of energy use. These estimates should be treated as order-of-magnitude values rather than precise measurements. Table 1 marks energy for the other two as not established. FuXi publishes no inference time, and the published IFS timing is for a newer ensemble set-up rather than the HRES configuration scored here (Rasp et al., 2024). Table 1 counts inference only, and including training changes the ranking. With the same reconstruction and the training costs published in the original papers, training took about 2,900 kWh for GraphCast (four weeks on 32 Cloud TPU v4 devices), about 62,000 kWh for Pangu-Weather (four networks, each 16 days on 192 V100s) and about 37 kWh for Keisler (2022) (5.5 days on one A100). Spread over one year of four forecasts a day, training is more than 99 per cent of each system's lifecycle energy. Lifecycle EASS is then about 33 for Keisler (2022), 0.45 for GraphCast and 0.02 for Pangu-Weather, which falls from second to last (Supplementary Section S1). Physical models carry equivalent one-off costs in tuning, spin-up and development, so the same accounting applies to them.

**Box 1** **How we combine skill and energy**

Each system is a point on a plane with energy on one axis and skill on the other (Figs. 1c, 2a). Within a class, the empirical Pareto front joins the systems that no other system beats on both. A single score hides that trade-off, so EASS is used for ranking only.

**Medium-range weather** Fig. 1c, d

**Skill: computed for all six systems**

Day-5 500 hPa geopotential against ERA5, from WeatherBench 2 (2020, global). GenCast leads at $\chi_h$ = 0.907; IFS HRES is 0.878.

**Energy: measured for none of them**

4 of 6 publish an inference time and hardware (GraphCast ≈ 0.0011 kWh per day-5 forecast, reconstructed, good to an order of magnitude).

FuXi publishes no timing. IFS HRES has a published timing, but for a newer ensemble set-up, so its energy is not established for the configuration scored here.

No CPMIP equivalent exists for weather models.

**Climate simulation** Fig. 2

**Skill: computed for 159 models**

Reichler & Kim $I^2$ from archived output, DJF 1979–1999 against NCEP/NCAR R1, across CMIP1–6.

**Energy: measured, but only from CMIP6**

CPMIP JPSY (Acosta et al. 2024) for 13 models; median 89 kWh per simulated year, spanning 7–464, a factor of 65.

CPMIP begins at CMIP6, so CMIP1–5 carry none.

AI emulators are not shown: no skill computed on this footing, and their energy is unmeasured.

**Combining the two: the Skill–Energy plane and EASS**

Skill: $\chi_h$ = 1 / (1 + ξ). ξ is the error relative to a reference: the forecast error over that of climatology for weather, and the Reichler & Kim index $I^2$ for climate.

Score: EASS = $\chi_h$ / E, in skill per kWh. Higher is better. It ranks systems within a class and is reported alongside the front, not instead of it.

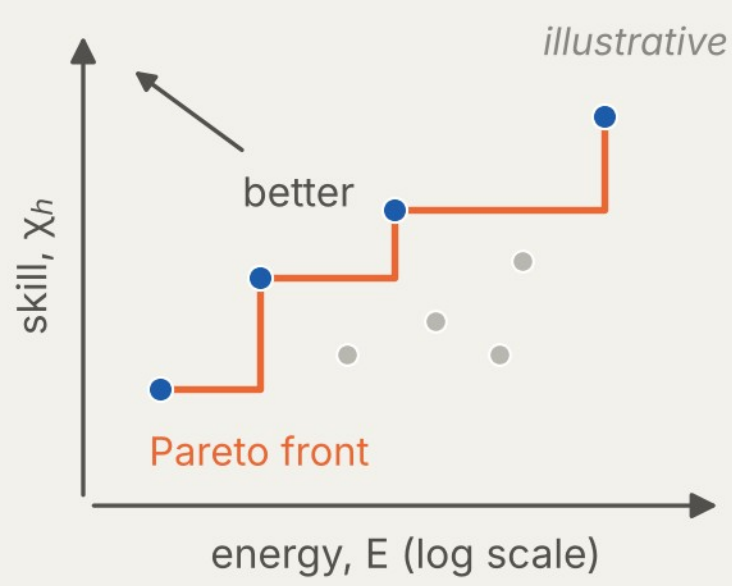

Weather and climate are never compared with each other, because their energy units (per day-5 forecast and per simulated year) and their skill targets differ. Skill is computed for every system here, but energy is measured for few of them.

## The Net-Zero Readiness Levels (NZRL)

**Box 2. Net-Zero Readiness Levels.**

We propose the Net-Zero Readiness Levels as a five-step pathway towards complete, transparent and independently verified reporting of energy use and emissions. Each level builds on the requirements below it and requires published evidence that a third party can check.

Level 1 (reported): Energy use for at least one production workflow is published, with the measurement tool and the accounting boundary stated.

Level 2 (standardised): Energy use is published in a common format and linked to the simulations it describes, enabling comparisons across models and centres. CPMIP provides this reporting framework.

Level 3 (linked to skill): Energy use is reported alongside a skill metric evaluated on the same runs, allowing readers to calculate a skill-per-energy score such as EASS.

Level 4 (contextualised): Reports also state the facility's power-usage effectiveness and the carbon intensity of its electricity supply, allowing comparisons between centres and machines.

Level 5 (complete and audited): Reports include amortised one-off costs and allocated emissions embodied in hardware, and the accounting is externally audited.

Supplementary Section S3 and Table S7 give the full requirements.

We used CPMIP data to assess how far the CMIP6 community has progressed along this reporting pathway (Box 2). Reporting energy in joules per simulated year through CPMIP meets Levels 1 and 2 when accompanied by the required measurement and accounting information (Acosta et al., 2024). Of the 30 institutions with a CMIP6 historical run evaluated here, nine (30%) reach Level 2. For the remaining 21 (70%), we found no published energy-per-simulated-year data in the evidence assessed. No institution reaches Level 3, because none reports a skill-per-energy score alongside conventional skill. We propose this as the next step: linking existing energy and skill assessments through a shared reporting convention. Consistent application of the pathway requires governance, shared evidence requirements and periodic review. The WCRP CMIP7 Energy and Carbon Footprint Task Team could coordinate this work. Figure 3 and Table S8 present the assessment.

## Four pillars of the Skill–Energy Frontier

Moving up the readiness scale makes energy use, and ultimately its carbon intensity, visible. Reducing it while preserving scientific value requires changes in how simulations are designed, executed and analysed. We identify four pillars: simulation design and reuse, emerging hardware, reduced data movement, and targeted AI (Fig. 4). The observed reduction in error of 13.5 per cent per doubling of estimated compute describes historical progress, not a fixed scientific limit. Across the thirteen CMIP6 models with CPMIP measurements, energy use per simulated year varies 65-fold (Fig. 2b; Supplementary Table S3 and Fig. S1). This spread reflects differences in models, hardware and execution, and cannot be attributed to computational efficiency alone. It still shows where energy demand might be reduced without compromising the scientific task.

**Pillar I: Simulation design and reuse**

The first pillar concerns the scientific value returned by a simulation relative to its energy demand. Intercomparisons from CMIP3 to CMIP6 have generated tens of petabytes of output, some of which remains under-analysed because extracting useful information requires substantial resources (Meehl et al., 2007; Taylor et al., 2012; Eyring et al., 2016). More output does not by itself bring more knowledge. Agreement with current observations is also not the only measure of a model's value. Diverse representations of physical processes remain important, and skill within the observed range does not guarantee performance in a changed climate (Knutti et al., 2010; Eyring et al., 2019). Although quantifying scientific value remains an open problem, explicit energy accounting makes the cost of different experimental choices visible.

This pillar has three practical consequences. First, experiments should address clearly defined questions, with their expected contribution weighed against their computational and energy costs (Juckes et al., 2020). Second, existing archives, emulators or AI surrogates should be considered before new runs are commissioned, provided they remain reliable for the question asked (Watson-Parris et al., 2022). Third, ensembles should be no larger than the scientific question requires. That size is rarely known in advance and should be treated as a research outcome (Meehl et al., 2014). Supplementary Section S4.3 discusses each consequence.

**Pillar II: Emerging hardware**

Hardware and software co-design offer opportunities to reduce energy demand, with increasing importance at higher resolution. A kilometre-scale Destination Earth configuration on MareNostrum 5 is estimated to require 10.3 MWh of node energy per simulated year, around 20 to 1,500 times the values for the CMIP6 models assessed here (Palomas et al., 2026). These systems deliver different capabilities, so the comparison illustrates the scale of demand rather than their relative efficiency.

Among the thirteen CMIP6 models with CPMIP measurements, energy use correlates weakly with resolution-derived compute estimates ($r=0.11$; Supplementary Fig. S1), but more strongly with core-hours per simulated year ($r=0.77$; Acosta et al., 2024, Table 3). Resolution alone therefore provides little explanation of the measured energy spread in this sample. Hardware, numerical implementation, model complexity and parallel execution all warrant closer investigation. The spread in carbon emissions is likely to be wider still, because these simulations ran on electricity grids of very different carbon intensity.

Many climate codes were written for central processing units (CPUs) and need substantial re-engineering to use accelerators efficiently. On graphics processing units (GPUs), ICON runs 5.5 times faster than on an optimised CPU baseline without changing its numerical schemes (Lapillonne et al., 2026), and a GPU-optimised 1 km configuration received the 2025 Gordon Bell Prize (Klocke et al., 2025).

Wafer-scale processors, which place hundreds of thousands of cores on a single wafer, suit kernels limited by memory bandwidth and communication. On one such processor, stencil computations ran fourteen times faster than on 128 conventional accelerators (Ozkan et al., 2025; Stawinoga et al., 2026).

Speed alone does not establish a benefit. Runtime, energy and emissions do not fall in the same proportion (Patterson et al., 2022; Henderson et al., 2020).

Accelerators also carry embodied emissions, for example 1,312 kg $CO_2e$ for one eight-GPU H100 baseboard (NVIDIA, 2024), and lifecycle results depend on utilisation and hardware lifetime (Golden et al., 2025). Hardware transitions should therefore be judged by energy and lifecycle emissions for a defined workload, not by speed or peak capacity alone (Supplementary Section S4.3).

**Pillar III: Reducing data movement**

Moving data can consume substantially more energy than performing arithmetic on it. Moving a unit of data across a chip, or off it, costs one to two orders of magnitude more energy than a floating-point operation. The cost depends on where the data reside and how far they must travel, increasing across the memory and network hierarchy (Horowitz, 2014; Unat et al., 2017; Shalf, 2020). Earth system science is particularly exposed because large datasets are routinely transferred to local systems for post-processing. Yet the energy attributable to these transfers is rarely quantified (Balaji et al., 2018; Petrie et al., 2021).

Analysing data near where they are stored can avoid repeated transfers, but the net benefit depends on the energy used by remote computation, storage and networking. Both workflows therefore need explicit accounting of transfers and storage (Lannelongue et al., 2021, 2023).

Compression reduces the data that must be stored and moved, and storage alone costs roughly 10 kg $CO_2e$ per terabyte per year (Lannelongue et al., 2021). Lossless compressors reduce climate datasets two- to four-fold without altering their values (Collet, 2018).

Error-bounded lossy methods such as ZFP and SZ achieve ten- to twenty-fold reductions for suitable data (Lindstrom, 2014; Di and Cappello, 2016), and many variables are stored with more precision than their information content warrants (Klöwer et al., 2021). Acceptable loss must be established for the intended analysis, particularly for extremes and trends. Neither kind of reduction implies a proportional saving in total workflow energy.

Neither cloud hosting nor local analysis is inherently the lower-carbon choice. The appropriate architecture minimises total energy or emissions for the required analysis, with the accounting boundary and any loss of precision stated (Supplementary Section S4.3).

**Pillar IV: Targeted AI and hybrids**

AI can reduce computational demand at several stages of the modelling workflow, from data preparation and simulation to post-processing and analysis. NeuralGCM, for example, demonstrates competitive performance on selected weather and climate evaluations at substantially lower runtime cost (Kochkov et al., 2024). Similarly, Antonio et al. (2026) demonstrate a skilful and stable mean climate when replacing a physical atmosphere with the ACE2 atmosphere emulator in the EC-Earth climate model, which reduces the computational cost by about 50 per cent compared with EC-Earth. Such results motivate targeted replacement or acceleration of expensive ESM components, provided their scientific performance is retained.

Climate applications impose requirements beyond short-range forecast accuracy. They demand stable long integrations, physically consistent conservation, credible low-frequency variability and reliable responses to changes in forcing outside the training range. Hybrid approaches can combine learned components with established physical structure, but their benefits must be tested against these requirements rather than inferred from inference speed alone.

Reuse is central to the potential energy advantage. Foundation models such as Aurora and Prithvi-WxC can spread pre-training costs across many downstream tasks (Bodnar et al., 2025; Schmude et al., 2024), but a lifecycle benefit follows only if downstream use offsets the pre-training investment (Ou et al., 2026).

The accounting boundary therefore extends beyond deployment to pre-training, architecture search, tuning and retraining, and to the simulations, reanalyses and data assimilation on which AI systems depend. Shared costs should be allocated across the applications that use them, without double counting (Wu et al., 2022; Supplementary Section S4.3).

The four pillars of the Skill–Energy Frontier reinforce one another. More selective experimental design reduces simulation demand and the volume of data requiring storage and analysis. Reuse can spread training and dataset-production costs across more applications. Hardware-aware implementations and reduced data movement can improve both physical and AI components. Each intervention should be assessed against the same criterion: whether it delivers more skill for a given energy budget, reduces energy demand at a given skill level, or achieves both. Improvements extend the empirical frontier when they outperform the best observed alternatives within the same comparison class and accounting boundary.

## Conclusions

We propose a change in how the Earth system modelling community measures progress: the Skill–Energy plane and its empirical Pareto front, the Energy-Adjusted Skill Score for ranking within a comparison class, and a five-level Net-Zero Readiness pathway based on published evidence. Adoption can begin on existing hardware through changes in measurement and reporting. The framework is a diagnostic. Applied without regard to scientific purpose, it could penalise expensive but necessary work, from high-resolution process studies to palaeoclimate experiments. Its aim is to make the energy cost of science visible while keeping scientific value first.

Three decades of coordinated climate modelling have established common protocols, experimental designs and evaluation standards that enable comparison across research groups and continents. The framework proposed here extends this tradition to energy use, which remains sparsely reported. The readiness pathway sets out the reporting steps, while the four pillars identify opportunities to reduce energy demand without compromising the scientific task.

The framework has limitations. EASS combines energy use with a single aggregated skill score and therefore inherits the strengths and weaknesses of that score. Assessments should also examine skill by variable, region and application, rather than relying on an aggregate alone. Alternative skill definitions can change EASS values and rankings, so sensitivity to the chosen metric should be reported. Future reporting should also include the carbon intensity of the electricity used, because the same energy use carries very different emissions on low- and high-carbon grids. Broader assessments should also report lifecycle emissions, including those embodied in hardware, separately from energy-based scores (NVIDIA, 2024; Golden et al., 2025; Palomas et al., 2026).

Practical barriers remain. Shared infrastructure complicates the allocation of energy to individual runs, and facility overhead varies across sites and over time. Reliable accounting may require integration with job schedulers, power monitoring and agreed methods for allocating shared costs. These practices are not yet consistently standardised. NZRL Levels 1 and 2 provide initial steps by requiring explicit accounting boundaries and common reporting formats, but implementation will require coordination between modelling centres and computing facilities.

Two broader risks also warrant attention. First, efficiency gains can produce rebound effects: cheaper forecasts or simulations may encourage enough additional use to offset, or even exceed, the energy saved per run (Kaack et al., 2022; Jiang et al., 2025; IEA, 2025). Efficiency alone is therefore not enough. Total energy use and emissions should be monitored alongside per-run measures and EASS.

Second, access to the infrastructure needed to train foundation weather models remains concentrated in a small number of organisations and countries, creating barriers for less-resourced research communities, including those in the Global South (Mozaffari et al., 2026). Open model weights, accessible training data and code, and support for adaptation and deployment can help distribute the benefits of these investments. These measures complement the reporting pathway: transparency about energy costs should be accompanied by efforts to broaden access to efficient prediction systems.

The first step is simple in principle, although it has a cost: report energy use alongside skill, with clear accounting boundaries and uncertainty estimates. Adoption by 2030 would establish a basis for assessing the scientific return on energy in future intercomparisons, including CMIP8.

**References**

Acosta, M. et al. The computational and energy cost of simulation and storage for climate science: lessons from CMIP6. Geosci. Model Dev. 17, 3081–3102 (2024).

Antonio, B., Strommen, K., Ortega, P. & Christensen, H. M. ACE2-NEMO: coupling an ML atmospheric emulator to a full-depth dynamical ocean model. Preprint at arXiv:2603.28704 (2026).

Balaji, V. et al. CPMIP: measurements of real computational performance of Earth system models in CMIP6. Geosci. Model Dev. 10, 19–34 (2017).

Balaji, V. et al. Requirements for a global data infrastructure in support of CMIP6. Geosci. Model Dev. 11, 3659–3680 (2018).

Balaji, V. Climbing down Charney's ladder: machine learning and the post-Dennard era of computational climate science. Philos. Trans. R. Soc. A 379, 20200085 (2021).

Balaji, V. et al. Are general circulation models obsolete? Proc. Natl Acad. Sci. USA 119, e2202075119 (2022).

Bauer, P., Thorpe, A. & Brunet, G. The quiet revolution of numerical weather prediction. Nature 525, 47–55 (2015).

Bauer, P., Dueben, P. D., Hoefler, T., Quintino, T., Schulthess, T. C. & Wedi, N. P. The digital revolution of Earth-system science. Nat. Comput. Sci. 1, 104–113 (2021).

Beucler, T., Pritchard, M., Yuval, J. et al. Climate-invariant machine learning. Sci. Adv. 10, eadj7250 (2024).

Beucler, T., Grundner, A., Shamekh, S., Ukkonen, P., Chantry, M. & Lagerquist, R. Distilling machine learning's added value: Pareto fronts in atmospheric applications. Artif. Intell. Earth Syst. 4, e240078 (2025).

Bi, K. et al. Accurate medium-range global weather forecasting with 3D neural networks. Nature 619, 533–538 (2023).

Bodnar, C. et al. A foundation model for the Earth System. Nature 641, 1180–1187 (2025).

Bracco, A. et al. Machine learning for the physics of climate. Nat. Rev. Phys. 7, 6–20 (2025).

Broude Geva, S., Chalker, A., Hillegas, C., Petravick, D., Sill, A. & Stewart, C. Results from a second longitudinal survey of academic research computing and data center usage. In Practice and Experience in Advanced Research Computing Conference (PEARC 2021). ACM. https://doi.org/10.1145/3437359.3465589 (2021).

Chen, K. et al. FengWu: pushing the skillful global medium-range weather forecast beyond 10 days lead. Preprint at arXiv:2304.02948 (2023a).

Chen, L. et al. FuXi: a cascade machine learning forecasting system for 15-day global weather forecast. npj Clim. Atmos. Sci. 6, 190 (2023b).

Cimadevilla, E. et al. ESGF virtual aggregation services for CMIP6 archives. Geosci. Model Dev. (2024).

Collet, Y. Zstandard compression and the application/zstd media type. RFC 8478, RFC Editor. https://doi.org/10.17487/RFC8478 (2018).

Courty, B. et al. Mlco2/codecarbon: V2.4.1. Zenodo. https://doi.org/10.5281/zenodo.11171501 (2024).

Dainelli, F., Mozaffari, A., Castaño, M., Gaya i Àvila, A., Palma Garcia, L., Melli, A., Dimdore-Miles, O. & Duarte, A. A checklist to assess the energy and carbon impacts of ML/AI applications in Earth system modeling. Preprint at arXiv:2609.00847 (2026).

Di, S. & Cappello, F. Fast error-bounded lossy HPC data compression with SZ. In 2016 IEEE International Parallel and Distributed Processing Symposium (IPDPS) 730–739. https://doi.org/10.1109/IPDPS.2016.11 (2016).

Eyring, V. et al. Overview of the Coupled Model Intercomparison Project Phase 6 (CMIP6). Geosci. Model Dev. 9, 1937–1958 (2016).

Eyring, V. et al. Taking climate model evaluation to the next level. Nat. Clim. Change 9, 102–110 (2019).

Eyring, V. et al. AI-empowered next-generation multiscale climate modelling for mitigation and adaptation. Nat. Geosci. 17, 963–971 (2024).

Golden, A. et al. Wafer-scale systems: a carbon perspective. ACM SIGENERGY Energy Inform. Rev. 5, 118–124 (2025).

Grundner, L. et al. Interpretable hybrid AI for cloud parameterization in climate models. Preprint at arXiv:2505.04358 (2025).

Henderson, P. et al. Towards the systematic reporting of the energy and carbon footprints of machine learning. J. Mach. Learn. Res. 21, 1–43 (2020).

Horowitz, M. Computing's energy problem (and what we can do about it). In 2014 IEEE International Solid-State Circuits Conference Digest of Technical Papers (ISSCC) 10–14. https://doi.org/10.1109/ISSCC.2014.6757323 (2014).

IEA. Energy and AI: energy demand from AI. IEA, Paris (2025).

Jiang, Z. et al. From efficiency gains to rebound effects: the problem of Jevons' paradox in AI's polarized environmental debate. In Proc. FAccT. ACM (2025).

Juckes, M. et al. The CMIP6 Data Request. Geosci. Model Dev. 13, 201–224 (2020).

Kaack, L. H., Donti, P. L., Strubell, E., Kamiya, G., Creutzig, F. & Rolnick, D. Aligning artificial intelligence with climate change mitigation. Nat. Clim. Change 12, 518–527 (2022).

Keisler, R. Forecasting global weather with graph neural networks. Preprint at arXiv:2202.07575 (2022).

Khandelwal, A., Yun, T., Nayak, N. V., Merullo, J., Bach, S. H., Sun, C. & Pavlick, E. $100K or 100 days: trade-offs when pre-training with academic resources. Preprint at arXiv:2410.23261 (2024).

Klocke, D. et al. Computing the full Earth system at 1 km resolution. In Proceedings of the International Conference for High Performance Computing, Networking, Storage and Analysis (SC25). ACM (2025).

Klöwer, M., Razinger, M., Dominguez, J. J., Düben, P. D. & Palmer, T. N. Compressing atmospheric data into its real information content. Nat. Comput. Sci. 1, 713–724 (2021).

Knutti, R., Furrer, R., Tebaldi, C., Cermak, J. & Meehl, G. A. Challenges in combining projections from multiple climate models. J. Clim. 23, 2739–2758 (2010).

Kochkov, D. et al. Neural general circulation models for weather and climate. Nature 628, 517–523 (2024).

Lam, R. et al. Learning skillful medium-range global weather forecasting (GraphCast). Science 382, eadi2336 (2023).

Lannelongue, L., Grealey, J. & Inouye, M. Green algorithms: quantifying the carbon footprint of computation. Adv. Sci. 8, 2100707 (2021).

Lannelongue, L., Aronson, H.-E. G., Bateman, A. et al. GREENER principles for environmentally sustainable computational science. Nat. Comput. Sci. 3, 514–521 (2023).

Lapillonne, X. et al. Operational numerical weather prediction with ICON on GPUs (version 2024.10). Geosci. Model Dev. 19, 755–772 (2026).

Lindstrom, P. Fixed-rate compressed floating-point arrays. IEEE Trans. Visual. Comput. Graphics 20, 2674–2683 (2014).

Manabe, S. & Stouffer, R. J. Two stable equilibria of a coupled ocean–atmosphere model. J. Climate 1, 841–866 (1988).

McCarthy, G. et al. Perspectives on the Atlantic Meridional Overturning Circulation. Earth Surf. Dyn. 11, 131–148 (2023).

Meehl, G. A. et al. Coupled Model Intercomparison Project. Bull. Am. Meteorol. Soc. 81, 313–318 (2000).

Meehl, G. A. et al. The WCRP CMIP3 multimodel dataset: a new era in climate change research. Bull. Am. Meteorol. Soc. 88, 1383–1394 (2007).

Meehl, G. A. et al. Decadal climate prediction: an update from the trenches. Bull. Am. Meteorol. Soc. 95, 243–267 (2014).

Molteni, F. et al. The ECMWF ensemble prediction system: methodology and validation. Q. J. R. Meteorol. Soc. 122, 73–119 (1996).

Moon, H. et al. Benefits of high-resolution climate models for precipitation. J. Climate 35, 3499–3514 (2022).

Mozaffari, A. et al. The rise of AI in weather and climate information and its impact on global inequality. npj Clim. Action https://doi.org/10.1038/s44168-026-00412-z (2026).

Neumann, P. et al. Assessing the scales in numerical weather and climate predictions: will exascale be the rescue? Philos. Trans. R. Soc. A 377, 20180148 (2019).

Nature Computational Science. The carbon footprint of computational research. Nat. Comput. Sci. 3, 659 (2023).

NVIDIA Corporation. PCF summary for NVIDIA HGX H100. NVIDIA product carbon footprint disclosure (2024).

Ou, Y. et al. Artificial intelligence to support cross-disciplinary climate change research. Nat. Clim. Change (2026).

Ozkan, M. et al. Performance, efficiency, and cost analysis of wafer-scale AI accelerators vs. single-chip GPUs. Preprint at ScienceDirect (2025).

Palmer, T. et al. The ECMWF ensemble prediction system: looking back (more than) 25 years and projecting forward 25 years. Q. J. R. Meteorol. Soc. 145, 12–24 (2019).

Palomas, S., Aparici, P., Utrera, G. & Acosta, M. C. Energy and $CO_2$ footprint of climate model intercomparison projects. Preprint at arXiv:2608.23509 (2026).

Pathak, J. et al. FourCastNet: a global data-driven high-resolution weather model using adaptive Fourier neural operators. Preprint at arXiv:2202.11214 (2022).

Patterson, D. et al. The carbon footprint of machine learning training will plateau, then shrink. Computer 55, 18–28 (2022).

Petrie, R. et al. Coordinating an operational data distribution network for CMIP6 data. Geosci. Model Dev. 14, 629–644 (2021).

Price, I. et al. Probabilistic weather forecasting with machine learning. Nature 637, 84–90 (2025).

Rasp, S. et al. WeatherBench 2: a benchmark for the next generation of data-driven global weather models. J. Adv. Model. Earth Syst. 16, e2023MS004019 (2024).

Reichler, T. & Kim, J. How well do coupled models simulate today's climate? Bull. Am. Meteorol. Soc. 89, 303–311 (2008).

Rieutord, T., Mignon, C., McAufield, E. & Gleeson, E. Energy and carbon footprint considerations for data-driven weather forecasting models. Weather, https://doi.org/10.1002/wea.70035 (2026).

Schmude, J. et al. Prithvi WxC: foundation model for weather and climate. Preprint at arXiv:2409.13598 (2024).

Schulthess, T. C., Bauer, P., Wedi, N., Fuhrer, O., Hoefler, T. & Schär, C. Reflecting on the goal and baseline for exascale computing: a roadmap based on weather and climate simulations. Comput. Sci. Eng. 21, 30–41 (2019).

Schwartz, R., Dodge, J., Smith, N. A. & Etzioni, O. Green AI. Commun. ACM 63, 54–63 (2020).

Shalf, J. The future of computing beyond Moore's Law. Philos. Trans. R. Soc. A 378, 20190061 (2020).

Stawinoga, N., Katz, D., Lydike, A., Zarins, J., Brown, N., Bisbas, G. & Grosser, T. An MLIR lowering pipeline for stencils at wafer-scale. In Proc. ASPLOS '26, Vol. 2, Article 1. ACM (2026).

Taylor, K. E., Stouffer, R. J. & Meehl, G. A. An overview of CMIP5 and the experiment design. Bull. Am. Meteorol. Soc. 93, 485–498 (2012).

Unat, D. et al. Trends in data locality abstractions for HPC systems. IEEE Trans. Parallel Distrib. Syst. 28, 3007–3020 (2017).

Watson-Parris, D. et al. ClimateBench v1.0. J. Adv. Model. Earth Syst. 14, e2021MS002954 (2022).

WCRP. CMIP7 Energy & Carbon Footprint Task Team Report. WCRP Rep. (2024).

Weijer, W. et al. AMOC in CMIP6. Geophys. Res. Lett. 47, e2020GL087786 (2020).

Wu, C. J. et al. Sustainable AI: environmental implications, challenges and opportunities. In Proc. Machine Learning and Systems Vol. 4, 795–813 (2022).

Xin, X. et al. East Asian summer monsoon in CMIP5 and CMIP6. Int. J. Climatol. 40, 6420–6438 (2020).

Table 1. Energy-adjusted skill for representative Earth system prediction systems. $\chi_h = 1/(1 + \xi)$ throughout. [NWP]: $\xi$ is the forecast mean squared error divided by that of climatology, for 500 hPa geopotential height at a 5-day lead against ERA5, global, 2020 (WeatherBench 2); E is kWh per day-5 forecast. [Climate]: $\xi$ is the Reichler & Kim index $I^2$, evaluated for DJF 1979-1999 against NCEP/NCAR R1 on sea-level pressure, surface air temperature and precipitation, normalised to the CMIP3 multi-model mean; E is kWh per simulated year from the CPMIP metric of Acosta et al. (2024). EASS = $\chi_h$ / E. The two classes use different skill targets and different energy denominators and are not compared. Values are operational inference costs; training is excluded (Supplementary Section S1). Where no inference time is published for the configuration scored, E is recorded as not established and no EASS is given. Per-variable skill decomposition is given in Tables S4 and S5.

| System | Class | Era | $\chi_h$ | Energy | EASS | Basis |
|---|---|---|---|---|---|---|
| CMIP3 (median model) | Climate | ~2005 | 0.529 | not measured | — | per simulated year |
| CMIP5 (median model) | Climate | ~2011 | 0.585 | not measured | — | per simulated year |
| CMIP6 (median model) | Climate | ~2019 | 0.622 | 89 kWh | 0.0070 | per simulated year |
| IFS HRES | NWP | 2023 | 0.878 | not established | — | per day-5 forecast |
| Keisler (2022) | NWP | 2022 | 0.849 | 0.000062 kWh | 13,600 | per day-5 forecast |
| Pangu-Weather | NWP | 2023 | 0.886 | 0.00041 kWh | 2,170 | per day-5 forecast |
| GraphCast | NWP | 2023 | 0.900 | 0.0011 kWh | 803 | per day-5 forecast |
| FuXi | NWP | 2023 | 0.898 | not established | — | per day-5 forecast |
| GenCast | NWP | 2024 | 0.907 | 0.30 kWh | 3.0 | per day-5 forecast |

## Figures

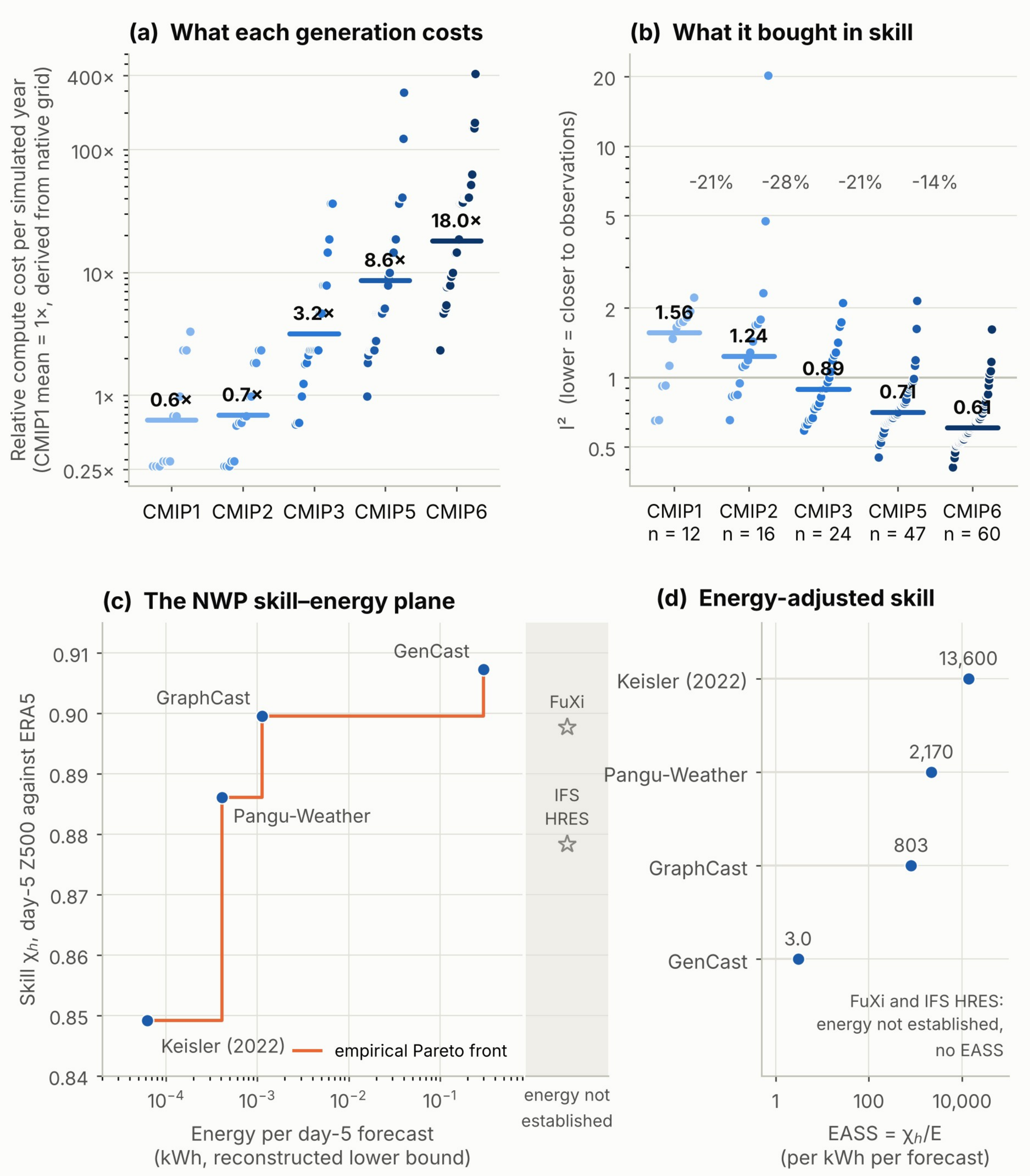

Figure 1. Compute and skill across CMIP generations (a, b), and medium-range weather forecasting on the Skill–Energy plane (c, d). (a) Relative computational scale per model, derived from each model's native horizontal grid and normalised so that the arithmetic mean of the CMIP1 models is 1. Bars are generation geometric means: 0.6x, 0.7x, 3.2x, 8.6x and 18.0x. This axis measures computational scale, not energy, and is a lower bound. (b) Reichler & Kim performance index $I^2$ for 159 models, lower values indicating closer agreement with observations: CMIP1 (n = 12), CMIP2 (16), CMIP3 (24), CMIP5 (47),

CMIP6 (60). Bars are generation medians: 1.56, 1.24, 0.89, 0.71 and 0.61, and percentages give the change at each step. Colours shade from light (CMIP1) to dark (CMIP6). (c) Medium-range numerical weather prediction (NWP): skill against energy per day-5 forecast for the four systems that publish an inference time and hardware, with the empirical Pareto front (orange staircase). Energy is reconstructed from the published timing and every value is a lower bound. Bars span utilisation of 60 to 80 per cent. Skill is $\chi_h = 1/(1 + \xi)$ for 500 hPa geopotential height at a 5-day lead against ERA5 (WeatherBench 2, 2020). The grey strip places FuXi and IFS HRES by skill alone, because their energy is not established for the configuration scored. (d) Energy-adjusted skill, EASS = $\chi_h/E$, for the same four systems, a within-class statistic not comparable with Fig. 2. GenCast is costed for all 50 ensemble members. Supplementary Section S4.4 gives the full method for each panel.

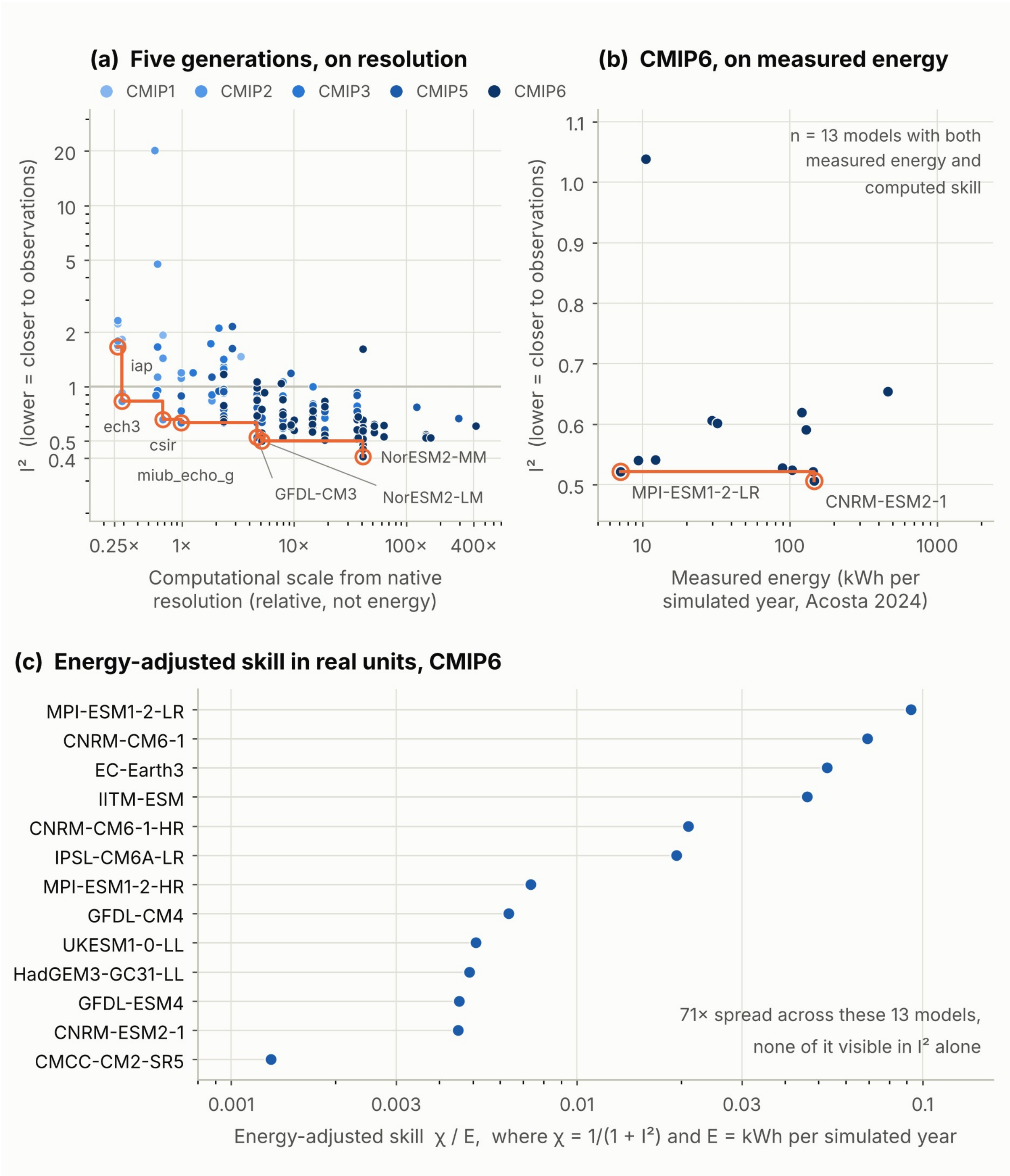

Figure 2. Climate simulation on the Skill–Energy plane. (a) The performance index $I^2$ of Fig. 1b, lower values indicating closer agreement with observations, for all five generations against the resolution-derived computational scale of Fig. 1a, with the empirical Pareto front (orange staircase). This axis is not energy: across the models for which both energy and resolution are available it correlates with measured energy at r = 0.11 on log-log axes (Supplementary Fig. S1). (b) $I^2$ for CMIP6 against measured energy,

with its own front. Energy is the CPMIP joules-per-simulated-year metric of Acosta et al. (2024), Table 3, converted to kWh per simulated year, for the $n = 13$ models that match a model scored here; where a model was measured at more than one centre the geometric mean is used. (c) EASS = $\chi_h/E$, with $\chi_h = 1/(1 + I^2)$, for those models, ranked, spanning a factor of 71. No measured energy exists for CMIP1 to CMIP5, because CPMIP reporting begins with CMIP6.

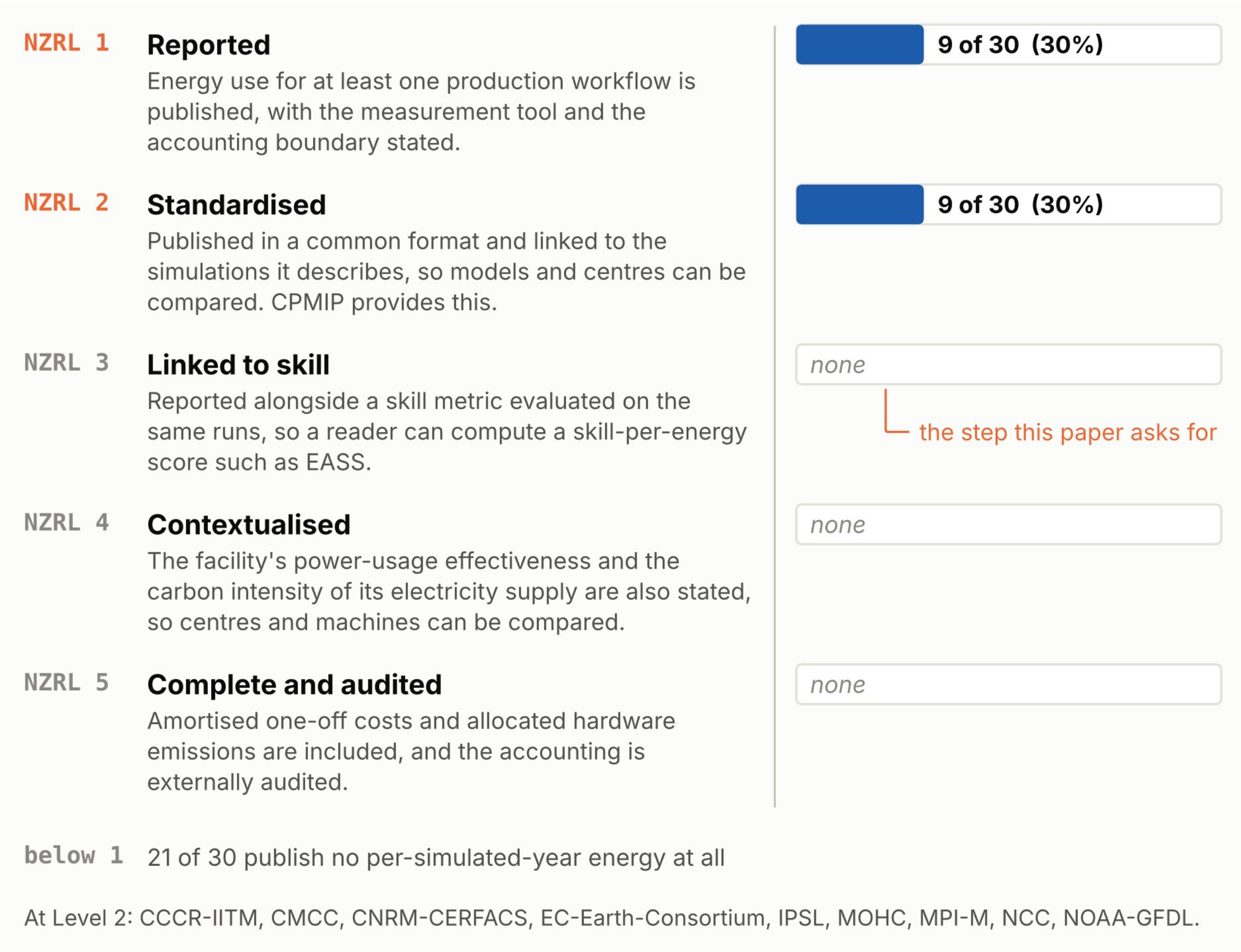

Figure 3. The readiness scale, and where the community sits on it. Each level requires a published artefact that a third party can verify, and the levels are cumulative (Box 2). Bars count institutions at or above each rung, over the 30 institutions holding a CMIP6 historical run scored in this paper. Nine reach level 2, all through CPMIP (Acosta et al., 2024). None reaches level 3, and 21 publish no energy per simulated year. The scale cannot separate absence of publication from absence of measurement, so its bottom boundary is open. Per-level requirements and the full scoring are in Supplementary Section S3 and Tables S7 and S8.

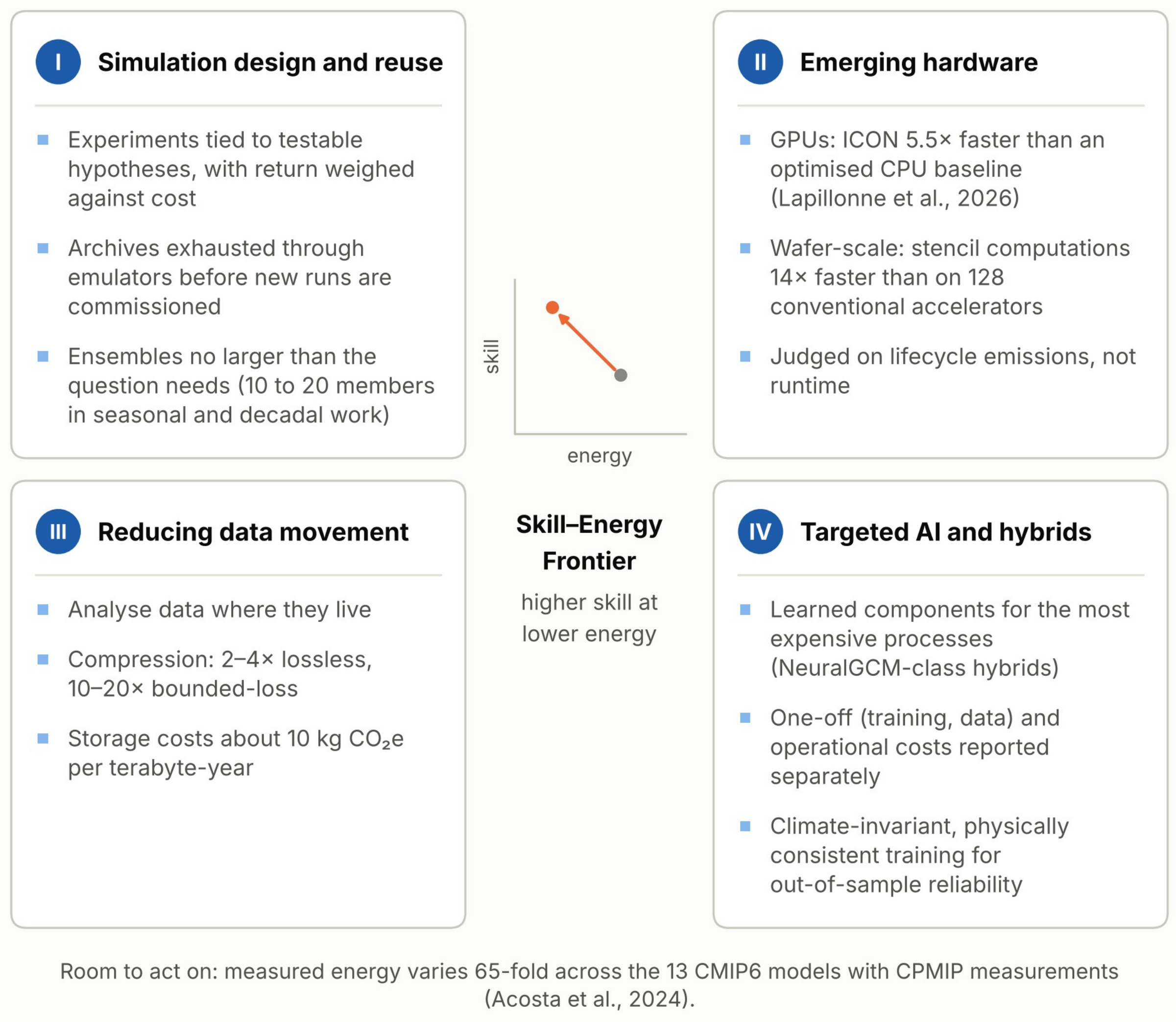

**Figure** 4. Four pillars for shifting the Skill–Energy Frontier. Each pillar acts on a different stage of the modelling workflow. I, simulation design and reuse: experiments tied to testable hypotheses, archives exhausted through emulators before new runs are commissioned, and ensembles no larger than the question needs. II, emerging hardware: GPUs, on which ICON runs 5.5 times faster than on an optimised CPU baseline (Lapillonne et al., 2026), and wafer-scale processors, on which stencil computations run fourteen times faster than on 128 conventional accelerators (Ozkan et al., 2025; Stawinoga et al., 2026). III, reducing data movement: analysis where the data live, and compression of two- to four-fold without loss and ten- to twenty-fold with bounded loss. IV, targeted AI and hybrids:

learned components for the most expensive processes, with one-off and operational costs reported separately. The pillars reinforce one another. They act on the 65-fold spread in measured energy across the 13 CMIP6 models with CPMIP measurements (Acosta et al., 2024).

# Supplementary Material

## Energy-Aware Earth System Science: A Community Roadmap

### Section S1. Energy and skill estimation for Table 1

**Skill metric mapping.** For [Climate] systems, the error ratio is the Reichler & Kim performance index $I^2$, which we computed from archived model output. $I^2$ is a normalised mean-squared-error index rather than a pattern correlation. For each model and variable, it is the area-weighted mean squared error of the simulated DJF climatology against observations, divided by the observed interannual variance. The normalised errors for each variable are then divided by the CMIP3 multi-model mean error for that variable before averaging, so that the CMIP3 multi-model mean $I^2$ is 1.0 by construction. The observational reference is NCEP/NCAR R1, for boreal winter (DJF) over 1979–1999, the satellite era. The variables are sea-level pressure, surface air temperature and precipitation. Only these three are used, because CMIP2 archives monthly fields of only these three and scoring one generation on more variables than another would make the comparison between generations meaningless. Our index follows the construction of Reichler & Kim (2008) but differs from theirs in three ways. We use the DJF climatology rather than the annual mean, three variables rather than fourteen, and NCEP/NCAR R1 as the single reference rather than the observational datasets and ERA-40 that they used. Absolute values are therefore not directly comparable with theirs, and for this reason we do not show their reanalysis value of 0.48 in Fig. 1b. Table S1 gives the computed generation medians. For [NWP] systems, skill is built in the same way. The error ratio is $\xi = MSE/MSE_{climatology}$ for 500 hPa geopotential height at a lead time of five days against ERA5, evaluated globally over 2020 from the published WeatherBench 2 benchmark (Rasp et al., 2024). It is mapped through the same bounded function, $\chi_h = 1/(1 + \xi)$. In both classes, the error is measured against a naive reference that uses no knowledge of the system: climatology for a forecast, and the observed interannual variance for a climate simulation. Table S2 gives the computed values. GenCast is an ensemble system and its ensemble-mean RMSE is used, which flatters it slightly relative to the deterministic entries. AIFS ENS and FengWu are not published in WeatherBench 2 and are therefore not scored. The two task classes measure different quantities, so scores in one class cannot be compared with scores in the other. Climate skill is assessed on climatological means rather than on individual forecasts, so the absolute values for CMIP entries are systematically lower. This is an artefact of the metric definitions and does not indicate lower predictive quality.

**Table S1.** Generation medians of the performance index and the compute scaling behind the rate of 13.5 per cent per doubling. $I^2$ is computed from archived output (DJF 1979–1999 against NCEP/NCAR R1, sea-level pressure, surface air temperature and precipitation, CMIP3 multi-model mean = 1.0). The compute scale is the generation geometric mean of the resolution-derived scale of Fig. 1a (arithmetic mean of the CMIP1 models = 1). The geometric means shown here lie below the arithmetic means, so CMIP1 is 0.63 rather than 1. Doublings are the base-2 logarithm of the compute ratio. The mean error reduction per doubling of the three defined steps is 13.5 per cent per doubling (standard deviation 0.7 points). The step from CMIP1 to CMIP2 bought only 0.14 doublings, so no rate is defined for it.

| Generation | Models scored | Median $I^2$ | Compute scale | Change in median $I^2$ | Compute ratio | Doublings | Error reduction per doubling (%) |
|---|---|---|---|---|---|---|---|
| CMIP1 | 12 | 1.560 | 0.63× | | | | |
| CMIP2 | 16 | 1.236 | 0.69× | −20.8% | 1.10 | 0.14 | not defined |
| CMIP3 | 24 | 0.892 | 3.18× | −27.8% | 4.60 | 2.20 | 12.64 |
| CMIP5 | 47 | 0.708 | 8.62× | −20.6% | 2.71 | 1.44 | 14.31 |
| CMIP6 | 60 | 0.607 | 18.03× | −14.3% | 2.09 | 1.06 | 13.45 |

**Table S2.** Medium-range weather systems: skill and the inputs to the reconstructed energy of Table 1. $\chi_h = 1/(1 + \xi)$, with $\xi$ the mean squared error of 500 hPa geopotential height at a five-day lead against ERA5, divided by that of climatology, global, 2020, from WeatherBench 2 (Rasp et al., 2024). Energy per day-5 forecast is device power × utilisation × devices × time, rescaled linearly to a five-day lead. The central value uses a utilisation of 0.7 and the range in brackets 0.6 to 0.8. Device power excludes host, network and cooling, so every value is a lower bound. EASS = $\chi_h/E$ in units of per kWh per forecast. Systems without a published inference time for the configuration scored are given no energy or EASS.

| System | Type | $\chi_h$ | Published inference time | Hardware | Members | Energy per day-5 forecast (kWh) | EASS | Source |
|---|---|---|---|---|---|---|---|---|
| GraphCast | AI | 0.900 | 10-day forecast in under 1 min | one Cloud TPU v4, 192 W | 1 | 0.0011 (0.00096–0.0013) | 803 | Lam et al. (2023) |
| Pangu-Weather | AI | 0.886 | 1.4 s per 24-h step, five steps to day 5 | one V100, 300 W | 1 | 0.00041 (0.00035–0.00047) | 2,170 | Bi et al. (2023) |
| GenCast | AI | 0.907 | about 8 min per member, 15-day forecast | one Cloud TPU v5, 192 W assumed | 50 | 0.30 (0.26–0.34) | 3.0 | Price et al. (2025) |
| FuXi | AI | 0.898 | not published | | | not established | | Chen et al. (2023b) |
| Keisler (2022) | AI | 0.849 | 0.04 s per 6-h step, 0.8 s to day 5 | one A100 40 GB, 400 W assumed | 1 | 0.000062 (0.000053–0.000071) | 13,600 | Keisler (2022) |
| IFS HRES | Physical | 0.878 | about 46 min per 15-day run on 64 nodes, newer ensemble set-up | | | not established | | ECMWF; Rasp et al. (2024) |

**Compute scaling.** The rate of 13.5 per cent per doubling quoted in the main text comes from the 159 models scored in Fig. 1b: the generation medians of $I^2$ above, and the generation geometric means of the resolution-derived compute scale of Fig. 1a over the same models. For each step between consecutive generations, we divide the percentage change in median $I^2$ by the number of compute doublings, the base-2 logarithm of the ratio of compute scales. Table S1 gives each step. The three steps give a mean of 13.5 per cent per doubling, with a standard deviation of 0.7 points. The step from CMIP1 to CMIP2 is excluded. It reduced $I^2$ by 20.8 per cent but bought only 0.14 doublings (a ratio of 1.10), so a rate per doubling is not well defined. At 13.5 per cent per doubling, halving the median error takes about 4.8 doublings, because 0.865 to the power 4.8 is 0.5. Since 2 to the power 4.8 is about 28, this is about 28 times the compute of CMIP6. The compute scale is derived from resolution and is not energy (Fig. S1).

**Energy estimation for CMIP6.** Energy is taken from the CPMIP column of joules per simulated year (JPSY) in Acosta et al. (2024), Table 3, converted to kWh per simulated year. These are measured values rather than estimates. Thirteen of the models reported there match models scored here. Table S3 lists them with their measured energy and skill. Where a model was measured at more than one centre, we use the geometric mean and report the spread. Measured energy across these thirteen models spans 7.1 to 464 kWh per simulated year, a factor of 65. No equivalent figure exists for CMIP1 to CMIP5, because CPMIP reporting began with CMIP6, so we give no measured energy for the earlier generations. Published core-hour totals (Balaji, 2021) describe hardware capability rather than the CMIP workload, and we do not use them as a proxy. We tested whether energy can be inferred from a model's published resolution, and found that it cannot. Across the 13 CMIP6 models with CPMIP measurements, resolution correlates with measured energy at $r = 0.26$, and a resolution-derived cost of the kind used in Figure 1a correlates at only $r = 0.11$. Core-hours per simulated year correlate at $r = 0.77$ (Acosta et al., 2024, Table 3; Fig. S1 shows the resolution comparison). Energy therefore has to be measured and published, because it cannot be reconstructed from a model description. We recommend that modelling groups report direct measurements, for example with CodeCarbon (Courty et al., 2024) or the Green Algorithms calculator (Lannelongue et al., 2021).

**Table S3.** The thirteen CMIP6 models with both measured energy and computed skill, ordered by energy. Energy is the CPMIP joules-per-simulated-year column of Acosta et al. (2024), Table 3, converted to kWh per simulated year. Where a model was measured at more than one centre, we use the geometric mean and give the spread factor. $I^2$ is the index of Table S1, $\chi_h = 1/(1 + I^2)$ and EASS = $\chi_h/E$ per kWh per simulated year (Fig. 2c).

| Model | CPMIP entries (centre: model) | Energy (kWh per simulated year) | Measurements | $I^2$ | $\chi_h$ | EASS |
|---|---|---|---|---|---|---|
| MPI-ESM1-2-LR | MPI:MPI-ESM1-LR | 7.11 | 1 | 0.522 | 0.657 | 0.092 |
| CNRM-CM6-1 | CNRM-CERFACS:CNRM-CM6-1 | 9.39 | 1 | 0.540 | 0.649 | 0.069 |
| IITM-ESM | IITM:IESM | 10.6 | 1 | 1.038 | 0.491 | 0.046 |
| EC-Earth3 | BSC:EC-Earth3 | 12.2 | 1 | 0.541 | 0.649 | 0.053 |
| CNRM-CM6-1-HR | CNRM-CERFACS:CNRM-CM6-1-HR | 29.7 | 1 | 0.606 | 0.623 | 0.021 |
| IPSL-CM6A-LR | IPSL:IPSL-CM6A | 32.2 | 1 | 0.602 | 0.624 | 0.019 |
| MPI-ESM1-2-HR | DKRZ:MPI-ESM1-HR | 89.2 | 1 | 0.528 | 0.654 | 0.0073 |
| GFDL-CM4 | GFDL:CM4-piC | 103 | 1 | 0.524 | 0.656 | 0.0064 |
| UKESM1-0-LL | NERC:UKESM1-0-LL, UKMO:UKESM1-0-LL | 121 | 2 (spread ×1.88) | 0.619 | 0.618 | 0.0051 |
| HadGEM3-GC31-LL | NERC:HadGEM3-GC3.1-LL, UKMO:HadGEM3-GC3.1-LL | 129 | 2 (spread ×1.15) | 0.591 | 0.629 | 0.0049 |
| GFDL-ESM4 | GFDL:ESM4-piC | 144 | 1 | 0.522 | 0.657 | 0.0046 |
| CNRM-ESM2-1 | CNRM-CERFACS:ESM2-1 | 147 | 1 | 0.507 | 0.664 | 0.0045 |
| CMCC-CM2-SR5 | CMCC:CM2-SR5 | 464 | 1 | 0.654 | 0.604 | 0.0013 |

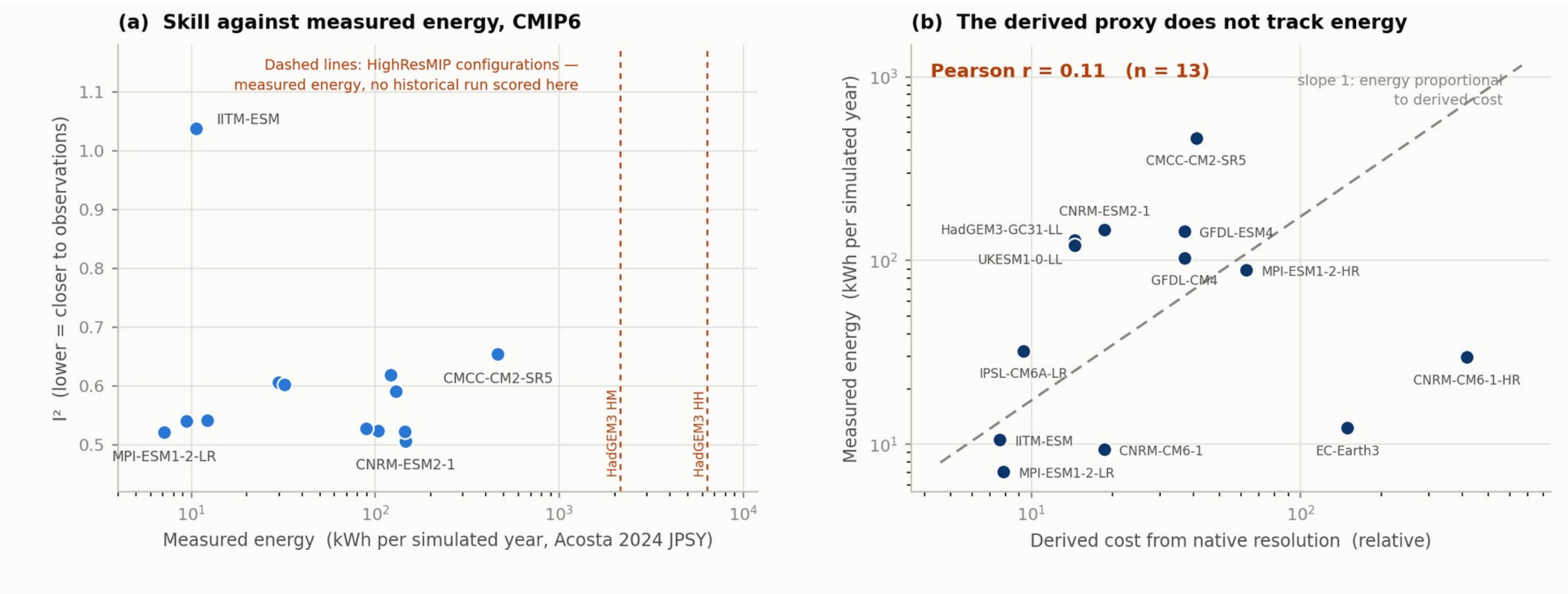

**Figure S1.** Measured energy for CMIP6, and why resolution cannot stand in for it. (a) The performance index I², lower values indicating closer agreement with observations, against measured energy for the 13 CMIP6 models that carry both. Energy is the CPMIP joules-per-simulated-year column of Acosta et al. (2024), Table 3, converted to kWh. Dashed lines mark two HighResMIP configurations (HadGEM3-GC31-HM and HH) that have measured energy but no historical run scored here. (b) Measured energy against the resolution-derived computational scale of Fig. 1a for the same 13 models. The two are effectively unrelated, with a Pearson correlation of r = 0.11 on log-log axes; the dashed line is slope 1, energy proportional to derived cost. Measured energy spans 7.1 to 464 kWh per simulated year, a factor of 65, while skill spans I² 0.507 to 1.038.

**Energy estimation for AI forecasters.** No operational weather system, physical or data-driven, publishes measured inference energy. The NWP energies in Table 1 are therefore reconstructed from each system's published inference time, the thermal design power (TDP) of the device it ran on, a utilisation fraction and the number of ensemble members charged. They are then rescaled from the lead time that the timing refers to, to the five-day lead at which skill is scored, assuming that cost grows linearly with lead time for autoregressive models and for the IFS. Device power is the TDP of the chip or board and excludes the host CPU, memory, storage and cooling, so every value is a lower bound. Table S2 gives the inputs for each system. Timings were checked against the primary papers on 15 October 2026. Four of the six systems publish an inference time: GraphCast (a 10-day forecast in under a minute on one Cloud TPU v4; Lam et al., 2023), Pangu-Weather (1.4 s per 24-hour step on one V100, five steps to day 5; Bi et al., 2023), Keisler (2022) (0.04 s per six-hour step, or 0.8 s to day 5, on one NVIDIA A100 after loading and compilation; Keisler, 2022, Section 3.2) and GenCast (about 8 minutes per 15-day ensemble member on one Cloud TPU v5; Price et al., 2025). No device power is published for the TPU v5, so we keep the TPU v4 value of 192 W. For Keisler (2022) we assume the 40 GB A100 SXM4 board at 400 W, as in Google Cloud A2 instances. The PCIe board (250 W) would lower its energy by 38 per cent and leave every ranking unchanged. WeatherBench 2 scores the forecasts of the model that Keisler (2022) describes, so the timing and the skill refer to the same model. GenCast is scored from its 50-member ensemble mean, so all 50 members are charged. Costing a single member would understate its energy fifty-fold and raise its EASS from 3.0 to about 150, still below Keisler (2022), Pangu-Weather and GraphCast in Table 1. FuXi (Chen et al., 2023b) publishes no inference time. For IFS HRES, WeatherBench 2 reports that a 15-day TCo1279 run in a newer operational-ensemble set-up takes about 52 minutes with I/O, or 46 minutes without, on 64 nodes of 128 AMD EPYC Rome cores each (Rasp et al., 2024). This is not the HRES configuration scored for 2020, so its energy is not established for the configuration evaluated here. Table 1 therefore gives no energy or EASS for FuXi or IFS HRES. All values reflect operational inference. They exclude one-off training, hyperparameter and architecture search, and the production of the reanalysis on which these models depend. Training costs are substantial. GraphCast required 28 days on 32 Cloud TPU v4 devices, and Pangu-Weather four networks, each trained for 16 days on 192 V100s, about 12,300 GPU-days. These costs must be amortised over the expected number of forecast cycles to obtain lifecycle EASS, which will be lower than the operational values in Table 1. Using the same reconstruction (device power × utilisation of 0.7 × devices × time), training comes to about 2,900 kWh for GraphCast, 62,000 kWh for Pangu-Weather and 37 kWh for Keisler (2022), which trained for 5.5 days on one A100 (Keisler, 2022, Section 3.3). Spread over one year of four forecasts a day (1,460 forecasts), this adds about 2.0, 42 and 0.025 kWh per forecast, more than 99 per cent of each system's lifecycle energy. Lifecycle EASS is then about 33 for Keisler (2022), 0.45 for GraphCast and 0.02 for Pangu-Weather, against 13,600, 803 and 2,170 at inference. GenCast publishes no training cost in a form we can reconstruct, so it is not ranked on this basis.

**Per-variable skill decomposition.** Table S4 reports per-variable $I^2$ and $\chi_h$ for CMIP3, CMIP5 and CMIP6 for Z500, T850, surface air temperature, precipitation and U850, together with the five-variable composite. The headline index of the main text averages three variables rather than five, as an arithmetic mean after normalisation by the CMIP3 multi-model mean. The decomposition matters because the aggregate index hides how uneven the progress has been. Table S5 gives the median error ratios of the three headline variables for all five generations. From CMIP1 to CMIP6, sea-level pressure improved 3.5-fold and surface air temperature 2.7-fold, while precipitation improved by 7 per cent. The dynamical fields therefore carry the headline improvement, and the hydrological field contributes little to it.

**Table S4.** Per-variable skill decomposition for CMIP3, CMIP5 and CMIP6. Each variable is normalised on the CMIP3 multi-model mean for that variable, so $I^2 = 1.0$ denotes the CMIP3 average for that field, and $\chi_h = 1/(1 + I^2)$. Entries are generation medians over the models that archive all five variables. The composite row is the median over models of each model's own five-variable index, not the mean of the five rows above it. Three archive defects are handled before normalisation. The observational file for 850 hPa temperature is in degrees Celsius while the models are in kelvin, and it is converted. CMIP3 ipsl_cm4 stores geopotential in place of geopotential height, although its files label the field as height in metres, so it is converted by dividing by standard gravity, g0 = 9.80665 m $s^{-2}$. CMIP6 E3SM-1-1 returns unmasked fill values and is excluded. The evaluation period and reference are as in Fig. 1b.

| Variable | CMIP3 $\chi_h$ | CMIP5 $\chi_h$ | CMIP6 $\chi_h$ | CMIP3 $I^2$ | CMIP5 $I^2$ | CMIP6 $I^2$ |
|---|---|---|---|---|---|---|
| Z500 (500 hPa geopotential) | 0.532 | 0.639 | 0.723 | 0.880 | 0.565 | 0.384 |
| T850 (850 hPa temperature) | 0.546 | 0.640 | 0.647 | 0.830 | 0.563 | 0.545 |
| Surface air temperature | 0.547 | 0.616 | 0.673 | 0.829 | 0.624 | 0.485 |
| Precipitation | 0.512 | 0.489 | 0.525 | 0.953 | 1.043 | 0.906 |
| U850 (850 hPa zonal wind) | 0.514 | 0.637 | 0.682 | 0.946 | 0.570 | 0.467 |
| Composite (five variables) | 0.525 | 0.589 | 0.630 | 0.903 | 0.697 | 0.587 |
| Models scored (n) | 23 | 47 | 59 | | | |

**Table S5.** Median error ratio of the three variables of the headline index, before normalisation on the CMIP3 multi-model mean: the mean squared error of the simulated DJF 1979–1999 climatology against NCEP/NCAR R1, divided by the observed interannual variance. Lower is better.

| Variable | CMIP1 | CMIP2 | CMIP3 | CMIP5 | CMIP6 | CMIP1 to CMIP6 |
|---|---|---|---|---|---|---|
| Sea-level pressure | 4.32 | 3.67 | 2.42 | 1.46 | 1.23 | 3.5-fold lower |
| Surface air temperature | 11.17 | 11.54 | 6.61 | 5.35 | 4.15 | 2.7-fold lower |
| Precipitation | 3.20 | 3.74 | 3.17 | 3.47 | 2.98 | 7 per cent lower |

**Sensitivity to the variable set.** The index uses three variables because CMIP2 archives only monthly fields of sea-level pressure, surface air temperature and precipitation. CMIP1 holds its upper-air fields as zonal cross-sections rather than maps. The five-variable set used from CMIP3 onwards therefore cannot be computed for either generation. This restriction is forced, but we tested it. CMIP3, CMIP5 and CMIP6 all archive the full set, so we computed both indices for these three generations, each normalised on its own CMIP3 multi-model mean. Table S6 gives the medians on both sets. The ordering and the sign of every step between generations are identical, so the restriction does not change the direction of the result. The size of the steps does change. They are larger on five variables (Table S6). The three-variable index understates the rate of improvement between generations by a factor of about 1.1, because the upper-air fields it omits improved faster than the surface ones. The headline result is therefore conservative. We found and handled three data defects in the upper-air fields during this check. The observational file for 850 hPa temperature is in degrees Celsius while the models are in kelvin. This scored every model at roughly 77,000 and pinned T850 at the reference mean, and it was corrected. CMIP3 ipsl_cm4 stores geopotential where it should store height, so it is converted by dividing by standard gravity. Its converted 500 hPa height has a global mean of 5,597 m against 5,637 m in NCEP/NCAR R1, and a DJF pattern correlation of 0.97 with the reanalysis. No other model in the three generations stores geopotential. CMIP6 E3SM-1-1 returns unmasked fill values and is excluded. These results come from 645 model-variable evaluations across 129 models.

**Table S6.** Median $I^2$ on three and on five variables for the generations that archive all five, each index normalised on its own CMIP3 multi-model mean, and the change in the median between generations.

| Variable set | CMIP3 | CMIP5 | CMIP6 | CMIP3 to CMIP5 | CMIP5 to CMIP6 |
|---|---|---|---|---|---|
| Three variables (psl, tas, pr) | 0.892 | 0.708 | 0.605 | −20.6% | −14.5% |
| Five variables (Z500, T850, tas, pr, U850) | 0.903 | 0.697 | 0.587 | −22.9% | −15.7% |

## Section S2. Definition of the skill score $\chi_h$

The skill component $\chi_h$ of the Energy-Adjusted Skill Score ($EASS = \chi_h/E$) is built in two steps, and it lies in the interval (0, 1].

**Step 1. Task-appropriate error relative to a naive reference.** For each prediction system, we form a dimensionless error ratio, $\xi \geq 0$, that compares the prediction error with that of a naive reference evaluated on the same set. The naive reference is climatology in both classes. For a forecast, it is the error made by predicting the climatological state. For a climate simulation, it is the observed interannual variance, which is the error made by predicting a randomly drawn year. For medium-range weather forecasts, $\xi$ is the mean squared error of the forecast divided by that of climatology, for 500 hPa geopotential height at a lead time of five days against ERA5, evaluated globally over 2020 from the published WeatherBench 2 benchmark evaluations:

$$\xi_{weather} = MSE_{forecast} \ / \ MSE_{climatology} \qquad (S2.1)$$

For climate simulations, $\xi$ is the Reichler & Kim performance index $I^2$, which we computed from archived model output. For each model and variable, it is the area-weighted mean squared error of the simulated DJF climatology against observations, divided by the observed interannual variance. The normalised errors for each variable are then divided by the CMIP3 multi-model mean error for that variable before averaging, so that the CMIP3 multi-model mean is 1.0 by construction:

$$\xi_{climate} = I^2 \qquad (S2.2)$$

The index is evaluated on sea-level pressure, surface air temperature and precipitation for boreal winter (DJF) over 1979–1999 against NCEP/NCAR R1. This is the only definition of climate skill used in this paper. Both classes share the construction of equation S2.3, and their error ratios differ only in what serves as the naive reference.

**Step 2. Bounded skill score.** The error ratio is mapped to a bounded skill score:

$$\chi_h = 1 / (1 + \xi) \qquad (S2.3)$$

**Properties.** This construction gives $\chi_h = 1$ for a perfect prediction ($\xi = 0$) and $\chi_h = 0.5$ for a prediction no better than the naive reference ($\xi = 1$). For predictions worse than the reference ($\xi > 1$), $\chi_h$ is below 0.5 and tends to 0. Forecast difficulty is absorbed into the reference rather than imposed through an explicit weighting for predictability. A more predictable field has a more skilful naive forecast, so exceeding it is correspondingly harder.

## Section S3. Net-Zero Readiness Levels: per-level requirements.

The scale has five cumulative levels. Each is defined by a published artefact, so that a third party can check where a centre sits. Table S7 sets out what each level requires. The levels are defined by publication rather than by internal practice, because only a published artefact can be verified from outside a centre. A readiness scale built on steps that cannot be verified would be a self-assessment. One boundary remains open by construction. The absence of a published figure cannot be distinguished from the absence of measurement, so the scale places a centre below level 1 without asserting that it measures nothing. Table S8 applies the scale to the CMIP6 institutions evaluated in this paper.

**Table S7.** Net-Zero Readiness Levels: what each level requires. The levels are cumulative, and each is defined by a published artefact that a third party can check.

| Level | Name | Requirement |
|---|---|---|
| NZRL 1 | Reported | The energy consumed by at least one production workflow is published in any form, with the measurement tool and the accounting boundary stated. |
| NZRL 2 | Standardised | Energy is published in a common form alongside the simulations it describes, so that it can be compared between models and centres. CPMIP (Acosta et al., 2024) meets this level and is the only such form currently in community use. |
| NZRL 3 | Linked to skill | Energy is published alongside a skill metric evaluated on the same runs, so that a reader can compute a skill-per-energy figure such as EASS. |
| NZRL 4 | Contextualised | The facility's power-usage effectiveness and the carbon intensity of its grid are declared with the energy, so that figures can be compared between centres and machines rather than describing one installation. |
| NZRL 5 | Complete and audited | Amortised one-off costs (training, dataset production and architecture search) and the emissions embodied in hardware are included, and the account is audited externally. |

**Table S8.** The Net-Zero Readiness scale applied to the 30 institutions that hold a CMIP6 historical run scored in this paper. Levels 1 and 2 are defined by publication. Reporting a value in joules per simulated year to CPMIP (Acosta et al., 2024, Table 3) therefore establishes level 2 by construction, because CPMIP is the standard form and Table 3 is the publication. Institutions absent from that table have no published energy per simulated year and are placed below level 1. The scale cannot distinguish the absence of publication from the absence of measurement, which is a limitation of any readiness scale built on what centres disclose. The CPMIP column names the reporting centre and model as printed in Table 3, which is not always the CMIP6 institution_id. Consortium models are run and metered by individual partners, and HadGEM3-GC3.1-LL and UKESM1-0-LL were each metered independently by two centres. No institution reaches level 3, which requires a skill-per-energy figure reported alongside conventional skill.

| Institution | NZRL | Models scored here | CPMIP Table 3 entries |
|---|---|---|---|
| AS-RCEC | <1 | TaiESM1 | |
| AWI | <1 | AWI-CM-1-1-MR; AWI-ESM-1-1-LR | |
| BCC | <1 | BCC-CSM2-MR; BCC-ESM1 | |
| CAMS | <1 | CAMS-CSM1-0 | |
| CAS | <1 | CAS-ESM2-0; FGOALS-f3-L; FGOALS-g3 | |
| CCCR-IITM | 2 | IITM-ESM | IITM:IESM |
| CCCma | <1 | CanESM5; CanESM5-CanOE | |
| CMCC | 2 | CMCC-CM2-HR4; CMCC-CM2-SR5; CMCC-ESM2 | CMCC:CM2-SR5 |

| Institution | NZRL | Models scored here | CPMIP Table 3 entries |
|---|---|---|---|
| CNRM-CERFACS | 2 | CNRM-CM6-1; CNRM-CM6-1-HR; CNRM-ESM2-1 | CNRM-CERFACS:CNRM-CM6-1-atm; CNRM-CERFACS:CNRM-CM6-1; CNRM-CERFACS:CNRM-CM6-1-HR-atm; CNRM-CERFACS:CNRM-CM6-1-HR; CNRM-CERFACS:ESM2-1-atm; CNRM-CERFACS:ESM2-1 |
| CSIRO | <1 | ACCESS-ESM1-5 | |
| CSIRO-ARCCSS | <1 | ACCESS-CM2 | |
| E3SM-Project | <1 | E3SM-1-0; E3SM-1-1; E3SM-1-1-ECA | |
| EC-Earth-Consortium | 2 | EC-Earth3; EC-Earth3-AerChem; EC-Earth3-CC; EC-Earth3-Veg-LR | BSC:EC-Earth3; BSC:EC-EarthVeg |
| FIO-QLNM | <1 | FIO-ESM-2-0 | |
| HAMMOZ-Consortium | <1 | MPI-ESM-1-2-HAM | |
| INM | <1 | INM-CM4-8; INM-CM5-0 | |
| IPSL | 2 | IPSL-CM6A-LR; IPSL-CM6A-LR-INCA | IPSL:IPSL-CM6A |
| KIOST | <1 | KIOST-ESM | |
| MIROC | <1 | MIROC-ES2L; MIROC6 | |
| MOHC | 2 | HadGEM3-GC31-LL; HadGEM3-GC31-MM; UKESM1-0-LL; UKESM1-1-LL | NERC:UKESM1-AMIP; NERC:UKESM1-0-LL; NERC:HadGEM3-GC3.1-LL; NERC:HadGEM3-GC3.1-HM; NERC:HadGEM3-GC3.1-HH; UKMO:HadGEM3-GC3.1-LL; UKMO:UKESM1-0-LL |
| MPI-M | 2 | MPI-ESM1-2-HR; MPI-ESM1-2-LR | DKRZ:MPI-ESM1-HR; MPI:MPI-ESM1-LR-ATM; MPI:MPI-ESM1-LR-LAND; MPI:MPI-ESM1-LR |
| MRI | <1 | MRI-ESM2-0 | |
| NASA-GISS | <1 | GISS-E2-1-G; GISS-E2-1-G-CC; GISS-E2-1-H; GISS-E2-2-G; GISS-E2-2-H | |
| NCAR | <1 | CESM2; CESM2-FV2; CESM2-WACCM; CESM2-WACCM-FV2 | |
| NCC | 2 | NorCPM1; NorESM2-LM; NorESM2-MM | NorESM:NorESM2-LM; NorESM:NorESM2-MM |
| NIMS-KMA | <1 | KACE-1-0-G; UKESM1-0-LL | |
| NOAA-GFDL | 2 | GFDL-CM4; GFDL-ESM4 | GFDL:OM4-p5; GFDL:ESM4-piC; GFDL:CM4-piC; GFDL:OM4-p25 |
| NUIST | <1 | NESM3 | |
| SNU | <1 | SAM0-UNICON | |
| THU | <1 | CIESM | |

## Section S4. Extended discussion from the main text

This section holds material that supports the main text but was moved here for length. The main text points to each part.

**S4.1 Existing benchmarks.** Existing benchmarks generally address predictive skill or computational cost separately. ClimateBench (Watson-Parris et al., 2022), ClimateBench-M, ClimSim-Online (Yu et al., 2025) and WxC-Bench (Shinde et al., 2024) evaluate AI climate components against physically grounded references. ClimateSet (Kaltenborn et al., 2023) standardises model output for comparison, and Luo et al. (2026) survey the wider field. These benchmarks do not integrate energy accounting into their skill assessments. In machine learning more broadly, MLPerf Power benchmarks energy efficiency alongside accuracy (Tschand et al., 2024), providing a precedent for joint evaluation. For Earth system models, CPMIP reports computational performance and energy for CMIP6 models (Acosta et al., 2024) but does not assess predictive skill.

**S4.2 Accounting levels.** The energy assigned to a system, and therefore its position on the plane and its EASS, depends on which costs are included. We define three accounting levels. Operational accounting counts the energy needed to produce one forecast or simulate a stated period, including data preparation, initialisation, simulation or inference, post-processing and storage (Table 1). Lifecycle accounting adds training, architecture search, dataset production and periodic retraining, spreading these costs across the forecasts or simulations produced over a stated period of use. System-level accounting also includes facility overhead, such as cooling, and an allocated share of the energy used to manufacture the hardware. Reports must state how shared costs are allocated and which components are measured, estimated or unavailable. A carbon-adjusted EASS includes the carbon intensity of the electricity grid. For AI systems, excluding training and other development costs can substantially overstate efficiency, particularly when a model is used infrequently or retrained often. The contribution of these costs depends on how widely and for how long the trained model is deployed. Training GraphCast, for example, took 28 days on 32 Cloud TPU v4 devices (Supplementary Section S1). This is the cost of training the final architecture. It excludes the iterative development that preceded it, and the same is true of the development and tuning of physical models, which are rarely costed. AI and hybrid systems should therefore report operational and lifecycle EASS separately.

**S4.3 The four pillars in detail.** The main text gives the main claim of each pillar. The full discussion follows.

**Pillar I, simulation design and reuse.** This pillar has three practical consequences. First, experiments should address clearly defined questions, with their expected contribution weighed against their computational and energy costs. Coordinated design can reduce duplication while preserving the replication needed to assess robustness across models (Juckes et al., 2020). Second, researchers should assess whether existing archives, emulators, statistical downscaling or AI surrogates can answer a question before commissioning new runs. For example, an emulator trained on climate-model output can reproduce temperature and precipitation responses across unseen scenarios at a fraction of the original computational cost (Watson-Parris et al., 2022). Such reuse requires checking that the surrogate remains reliable for the variables, scenarios and scales of interest. Third, ensembles should be no larger than needed to estimate uncertainty reliably for the scientific question being asked. This size is rarely known in advance and should be treated as a research outcome. Additional members are valuable when they improve estimates of forced responses, internal variability or rare events, but their marginal contribution should be assessed rather than assumed. Seasonal and decadal prediction studies illustrate how ensemble design affects this balance (Meehl et al., 2014). There is no universally sufficient ensemble size: the appropriate number depends on the model, variable, lead time and statistical quantity being estimated. Identifying when further members add little decision-relevant information is itself a research task.

**Pillar II, emerging hardware.** Many climate codes were developed for central processing units (CPUs) and require substantial re-engineering to use accelerators efficiently. Graphics processing units (GPUs) provide one route because they can process large arrays in parallel. At MeteoSwiss, ICON runs 5.5 times faster on GPUs than on an optimised CPU baseline in a socket-to-socket comparison, without changing its numerical schemes (Lapillonne et al., 2026). At 1 km resolution, a GPU-optimised ICON configuration sustained 146 simulated days per day on 8,000 accelerators and received the 2025 Gordon Bell Prize (Klocke et al., 2025).

Wafer-scale processors provide another route. Devices such as the Cerebras WSE-3 place hundreds of thousands of cores on a single silicon wafer, reducing reliance on communication between separate chips. This architecture may benefit weather and climate kernels limited by memory bandwidth and communication rather than arithmetic throughput. In recent benchmarks, stencil computations, central to many atmospheric numerical schemes, ran fourteen times faster on a single wafer-scale processor than on 128 conventional accelerators (Ozkan et al., 2025; Stawinoga et al., 2026).

Faster hardware brings two accounting challenges. First, runtime, energy and carbon emissions do not fall in the same proportion. Their relationship depends on power demand, utilisation, memory traffic, facility overhead and electricity carbon intensity (Patterson et al., 2022; Henderson et al., 2020). Reporting runtime alone therefore cannot establish an energy or emissions benefit.

Second, accelerators carry embodied emissions that must be considered alongside operational savings. NVIDIA reports 1,312 kg CO2-equivalent (CO2e) from manufacturing an eight-GPU HGX H100 baseboard, with 42 per cent attributed to high-bandwidth memory (NVIDIA, 2024). Comparisons of wafer-scale systems and GPU clusters likewise show that lifecycle results depend strongly on utilisation, hardware lifetime and the workload delivered (Golden et al., 2025). Resource-use measurements can help identify opportunities for improvement, but different measures of utilisation are not interchangeable. For example, the finding that more than half of GPU jobs on the National Energy Research Scientific Computing Center (NERSC)'s Perlmutter system use a quarter or less of available GPU memory does not, by itself, establish low computational utilisation (Li et al., 2023). Hardware transitions should be judged by energy and lifecycle emissions for a defined workload, not by speed or peak capacity alone.

**Pillar III, reducing data movement.** Analysing data near their storage location can reduce repeated transfers, but the net benefit depends on the full workflow. It includes the energy used by remote computation, storage, networking and any replicated data. Tools such as Green Algorithms can support estimates of computational footprints, while the GREENER principles provide guidance for reducing them (Lannelongue et al., 2021, 2023). Direct comparisons of near-data analysis and download-and-analyse workflows also require explicit accounting for transfers and storage, which are not captured by runtime alone.

Compression offers a complementary approach by reducing the data that must be stored and moved (Lindstrom, 2014). Its energy benefit depends on the balance between reduced input/output and the computation needed for compression and decompression. Storage itself costs roughly 10 kg CO2e per terabyte per year (Lannelongue et al., 2021). Lossless compressors can reduce climate datasets two- to four-fold, depending on the data and configuration, without altering their values (Collet, 2018; Blosc et al., 2023). These reductions can lower storage and transfer demand, but do not imply proportional savings in total workflow energy.

Lossy methods achieve larger reductions by discarding information within specified tolerances. ZFP and the error-bounded SZ compressor can achieve ten- to twenty-fold reductions for suitable datasets and error thresholds (Lindstrom, 2014; Di and Cappello, 2016). Bit grooming in the netCDF Operators has demonstrated storage savings of 25 to 80 per cent while preserving the precision required for the evaluated applications (Zender, 2016). Analyses of ERA5 and CMIP data also suggest that many variables are stored with substantially more precision than their information content warrants (Klöwer et al., 2021). Acceptable compression must nevertheless be established for the intended analysis, particularly for extremes, trends and derived quantities.

Neither cloud hosting nor local analysis is inherently the lower-carbon choice. The result depends on electricity supply, facility efficiency, access patterns, storage class and replication. A workflow that downloads an archive once and repeatedly analyses it locally may outperform one that streams the same data on every access. Conversely, shared analysis near an archive may avoid many redundant downloads. The appropriate architecture minimises total energy or emissions for the required analysis, with the accounting boundary and any loss of numerical precision stated explicitly.

**Pillar IV, targeted AI and hybrids.** Reuse is central to the potential energy advantage. Foundation models such as Aurora and Prithvi-WxC are pre-trained for adaptation to multiple downstream applications, potentially reducing the training required for each new task (Bodnar et al., 2025; Schmude et al., 2024). These savings depend on how often the model is reused, the cost of adaptation and the alternatives it replaces. A large pre-training investment does not establish a lifecycle benefit unless sufficient downstream use offsets that investment (Ou et al., 2026).

The relevant accounting boundary therefore extends beyond deployment. It includes pre-training, architecture search, tuning, maintenance and retraining, with shared costs allocated across the applications that use them (Wu et al., 2022). AI systems may also depend on high-resolution simulations, ensembles or reanalyses produced by conventional models. These dependencies should be declared, and their allocated costs reported without double counting. Data assimilation adds further demands in initialised prediction, seasonal forecasting and reanalysis production (Huang et al., 2024; Manshausen et al., 2024; Qu et al., 2024). Its costs should be reported separately or apportioned across the products and applications that depend on it.

**S4.4 Figure 1 methods.** Panel (a). The relative computational scale of each model is derived from its native horizontal grid as (grid points)^1.5, the scaling implied by the CFL timestep constraint, and normalised so that the arithmetic mean of the CMIP1 models is 1. The bars are generation geometric means, which lie below the arithmetic means, so CMIP1 sits at 0.6x rather than 1x. Vertical levels and added Earth-system complexity are not counted, so the growth is a lower bound. The axis measures computational scale and not energy (Fig. S1).

Panel (b). The Reichler & Kim performance index $I^2$ is computed from archived model output for boreal winter (DJF) 1979–1999 against NCEP/NCAR R1, averaged over sea-level pressure, surface air temperature and precipitation, and normalised so that the CMIP3 multi-model mean equals 1.0 by construction (Section S1). Two models whose archived record is shorter than 30 months, one in CMIP2 and one in CMIP6, are excluded and not shown.

Panel (c). The empirical Pareto front is the set of systems that no other system beats on both axes. Energy is reconstructed from the published timing and rescaled to a common day-5 horizon. The bars span utilisation of 60 to 80 per cent, and all values exclude host, network and cooling (Section S1 and Table S2). $\xi$ is the forecast mean squared error divided by that of climatology, for 500 hPa geopotential height at a 5-day lead against ERA5, global, 2020, from the WeatherBench 2 benchmark evaluations (Eqs. S2.1, S2.3). FuXi publishes no inference time, and the published IFS timing is for a newer ensemble set-up, so neither is given an energy.

Panel (d). EASS = $\chi h/E$ is a within-class statistic and is not comparable with Fig. 2. GenCast is costed for all 50 members of the ensemble on which it is scored.

## Supplementary references

References cited only in this Supplementary Material. All others are listed in the main text.

Blosc Project, Alted, F., Iborra, M., Guiñón, O., Ibáñez, D. & Barrachina, S. Using Blosc2 NDim as a fast explorer of the Milky Way (or any other NDim dataset). https://doi.org/10.25080/gerudo-f2bc6f59-000 (2023).

Huang, L. et al. DiffDA: a diffusion model for weather-scale data assimilation. Preprint at arXiv:2401.05932 (2024).

Kaltenborn, J. et al. ClimateSet: a large-scale climate model dataset for machine learning. In Advances in Neural Information Processing Systems 36 (NeurIPS 2023), Datasets and Benchmarks Track (2023).

Li, J., Bouhali, O., Dholakia, R., Doraiswamy, S. & Khan, S. M. Analyzing resource utilization in an HPC system: a case study of NERSC's Perlmutter. Preprint at arXiv:2301.05145 (2023).

Luo, J.-J. et al. AI for atmosphere–ocean sciences: advancements, challenges and ways forward. Natl. Sci. Rev. 13, nwag063 (2026).

Manshausen, P. et al. Generative data assimilation of sparse weather station observations at kilometre scales. Preprint at arXiv:2406.16947 (2024).

Qu, Y. et al. Deep generative data assimilation in multimodal setting. In Proc. CVPR (2024).

Shinde, R. et al. WxC-Bench: a novel dataset for weather and climate downstream tasks. Preprint at arXiv:2412.02780 (2024).

Tschand, A. et al. MLPerf Power: benchmarking the energy efficiency of machine learning systems from microwatts to megawatts for sustainable AI. Preprint at arXiv:2410.12032 (2024).

Yu, S. et al. ClimSim-Online: a large multi-scale dataset and framework for hybrid physics-ML climate emulation. J. Mach. Learn. Res. 26 (2025).

Zender, C. S. Bit grooming: statistically accurate precision-preserving quantization with compression, evaluated in the netCDF Operators (NCO, v4.4.8+). Geosci. Model Dev. 9, 3199–3211 (2016).